\documentclass[11pt, onecolumn, final]{IEEEtran}
\usepackage{amsmath, mathrsfs,amsfonts}
\usepackage{amsfonts}
\usepackage{amssymb}
\usepackage{acronym}
\usepackage{graphicx}

\def\ignore#1{}

\def\tl{\tilde}

\def\bs{\boldsymbol}
\def\R{\mathbb{R}}
\def\foral{\textrm{for all} \ }

\newtheorem{assumption}{Assumption}

\newtheorem{definition}{Definition}

\newtheorem{theorem}{Theorem}

\newtheorem{lemma}{Lemma}

               {\begin{list}{}{\leftmargin#1\rightmargin#2\topsep#3}\item{}}%
               {\end{list}}

\begin{document}
\title{On Resource Allocation in Fading Multiple Access Channels - An Efficient Approximate Projection Approach}



\author{Ali ParandehGheibi\thanks{A.\ ParandehGheibi is with the Laboratory for
Information and Decision Systems, Electrical Engineering and Computer Science Department,
Massachusetts Institute of Technology, Cambridge MA, 02139 (e-mail: parandeh@mit.edu)}, Atilla
Eryilmaz\thanks{A.\ Eryilmaz is with the Electrical and Computer Engineering, Ohio State
University, OH, 43210 (e-mail: eryilmaz@ece.osu.edu)}, Asuman Ozdaglar, and Muriel M\'edard\thanks{
A.\ Ozdaglar and M.\ M\'edard are with the Laboratory for Information and Decision Systems,
Electrical Engineering and Computer Science Department, Massachusetts Institute of Technology,
Cambridge MA, 02139 (e-mails: asuman@mit.edu, medard@mit.edu)}}

\markboth{LIDS Report 2787}{}
%

\maketitle

\thispagestyle{headings}

\begin{abstract}
We consider the problem of rate and power allocation in a multiple-access channel. Our objective is
to obtain rate and power allocation policies that maximize a general concave utility function of
average transmission rates on the information theoretic capacity region of the multiple-access
channel. Our policies does not require queue-length information. We consider several different
scenarios. First, we address the utility maximization problem in a non-fading channel to obtain the
optimal operating rates, and present an iterative gradient projection algorithm that uses
approximate projection. By exploiting the polymatroid structure of the capacity region, we show
that the approximate projection can be implemented in time polynomial in the number of users.
Second, we consider resource allocation in a fading channel. Optimal rate and power allocation
policies are presented for the case that power control is possible and channel statistics are
available. For the case that transmission power is fixed and channel statistics are unknown, we
propose a greedy rate allocation policy and provide bounds on the performance difference of this
policy and the optimal policy in terms of channel variations and structure of the utility function.
We present numerical results that demonstrate superior convergence rate performance for the greedy
policy compared to queue-length based policies. In order to reduce the computational complexity of
the greedy policy, we present approximate rate allocation policies which track the greedy policy
within a certain neighborhood that is characterized in terms of the speed of fading.
\end{abstract}


\section{Introduction}

Dynamic allocation of communication resources such as bandwidth or transmission power is a central
issue in multiple access channels in view of the time varying nature of the channel and the
interference effects. Most of the existing literature focuses on specific communication schemes
such as TDMA (time-division multiple access) \cite{TDMA}, CDMA (code-division multiple access)
\cite{CDMA1,CDMA3}, and OFDM (Orthogonal Frequency Division Multiplexing) \cite{berry07} systems.
An exception is the work by Tse \emph{et al.} \cite{Tse}, which consider the notion of
\emph{throughput capacity} for the fading channel with Channel State Information (CSI). The
throughput capacity is the notion of Shannon capacity applied to the fading channel, where the
codeword length can be arbitrarily long to average over the fading of the channel. Tse \emph{et
al.} \cite{Tse} consider allocation of rate and power to maximize a linear utility function of the
transmission rates over the throughput region, which characterizes the points on the boundary of
the throughput capacity region.

In this paper, we consider the problem of rate and power allocation in a multiple access channel
with perfect CSI. Contrary to the linear case in \cite{Tse}, we consider maximizing a general
utility function of transmission rates over the throughput capacity region. Such a general concave
utility function allows us to capture different performance metrics such as fairness or delay (cf.
Shenker \cite{She95}, Srikant \cite{Srikant}). Our contributions can be summarized as follows.

We first consider a non-fading multiple-access channel where we introduce a gradient projection
algorithm for the problem of maximizing a concave utility function of transmission rates over the
capacity region. We establish the convergence of the method to the optimal rate allocation. Since
the capacity region of the multiple-access channel is described by a number of constraints
exponential in the number of users, the projection operation used in the method can be
computationally expensive. To reduce the computational complexity, we introduce a new method that
utilizes \emph{approximate projections}. By exploiting the polymatroid structure of the capacity
region, we show that the approximate projection operation can be implemented in time polynomial in
number of users by using submodular function minimization algorithms. Moreover, we present a more
efficient algorithm for the approximate projection problem which relies on rate-splitting
\cite{Urbanke}. This algorithm also provides the extra information that allows the receiver to
decode the message by successive cancelation.

Second, we consider a fading multiple access channel and study the case where channel statistics
are known and transmission power can be controlled at the transmitters. Owing to strict convexity
properties of the capacity region along the boundary, we show that the resource allocation problem
for a general concave utility is equivalent to another problem with a linear utility. Hence, the
\emph{optimal} resource allocation policies are obtained by applying the results in \cite{Tse} for
the linear utility. Given a general utility function, the conditional gradient method is used to
obtain the corresponding linear utility.

If the transmitters do not have the power control feature and channel statistics are not known, the
throughput capacity region is a polyhedron and the strictly convexity properties of the region do
not hold any more. Hence, the previous approach is not applicable. In this case, we consider a
greedy policy, which maximizes the utility function for any given channel state. This policy is
suboptimal, however, we can bound the performance difference between the optimal and the greedy
policies. We show that this bound is tight in the sense that it goes to zero either as the utility
function tends to a linear function of the rates or as the channel variations vanish.

The greedy policy requires exact solution of a nonlinear program in each time slot, which makes it
computationally intractable. To alleviate this problem, we present approximate rate allocation
policies based on the gradient projection method with approximate projection and study its tracking
capabilities when the channel conditions vary over time. In our algorithm, the solution is updated
in every time slot in a direction to increase the utility function at that time slot. But, since
the channel may vary between time-slots, the level of these temporal channel variations become
critical to the performance. We explicitly quantify the impact of the speed of fading on the
performance of the policy, both for the worst-case and the average speed of fading. Our results
also capture the effect of the degree of concavity of the utility functions on the average
performance.

An important literature relevant to our work appears in the context of cross-layer design, where
joint scheduling-routing-flow control algorithms have been proposed and shown to achieve utility
maximization for concave utility functions while guaranteeing network stability (e.g.
\cite{linshr05, erysri05, neemodli05, sto05}). The common idea behind these schemes is to use
properly maintained queues to make dynamic decisions about new packet generation as well as rate
allocation.

Some of these works (\cite{erysri05, neemodli05}) explicitly address the fading channel conditions,
and show that the associated policies can achieve rates arbitrarily close to the optimal based on a
design parameter choice. However, the rate allocation with these schemes requires that a large
optimization problem requiring global queue-length information be solved over a complex feasible
set in every time slot. Clearly, this may not always be possible owing to the limitations of the
available information, the processing power, or the complexity intrinsic to the feasible set.
Requirement for queue-length information may impose much more overhead on the system than channel
state information. On the other hand, even in the absence of fading, the interference constraints
among nearby nodes' transmissions may make the feasible set so complex that the optimal rate
allocation problem becomes NP-hard (see \cite{eryozdmod07}). Moreover, the convergence results of
queue-length based policies (\cite{erysri05, neemodli05}) are asymptotic, and our simulation
results show that such policies may suffer from poor convergence rate. In fact, duration of a
communication session may not be sufficient for these algorithms to approach the optimal solution
while suboptimal policies such as the greedy policy seems to have superior performance when
communication time is limited, even though the greedy policy does not use queue-length information.

In the absence of fading, several works have proposed and analyzed approximate randomized and/or
distributed rate allocation algorithms for various interference models to reduce the computational
of the centralized optimization problem of the rate allocation policy (\cite{tas98, linshr05,
modshazus06, eryozdmod07, sanbuisri07, joolinshr07}). The effect of these algorithms on the utility
achieved is investigated in \cite{eryozdmod07,eryozdshamod07}. However, no similar work exists for
fading channel conditions, where the changes in the fading conditions coupled with the inability to
solve the optimization problem instantaneously make the solution much more challenging.

Other than the papers cited above, our work is also related to the work of Vishwanath \emph{et al.}
\cite{Vishwanath} which builds on \cite{Tse} and takes a similar approach to the resource
allocation problem for linear utility functions. Other works address different criteria for
resource allocation including minimizing delay by a queue-length based approach \cite{yeh},
minimizing the weighted sum of transmission powers \cite{power_min}, and considering Quality of
Service (QoS) constraints \cite{QoS}. In contrast to this literature, we consider the utility
maximization framework for general concave utility functions.

The remainder of this paper is organized as follows: In Section II, we introduce the model and
describe the capacity region of a fading multiple-access channel.  In Section III, we consider the
utility maximization problem in a non-fading channel and present the gradient projection method
with approximate projection. In Section IV, we address the resource allocation problem with power
control and known channel statistics. In Section V, we consider the same problem without power
control and knowledge of channel statistics. We present the greedy policy and approximate rate
allocation policies and study their tracking behavior. Section VI provides the simulation results,
and we give our concluding remarks in Section VII.

Regarding the notation, we denote by $x_i$ the $i$-th component of a vector $\bs x$. We denote the
nonnegative orthant by $\mathbb{R}^n_+$, i.e., $\mathbb{R}^n_+ = \{\bs x\in \mathbb{R}^n\mid \bs
x\ge 0\}$. We write $\bs x'$ to denote the transpose of a vector $\bs x$. We use the notation
$\textbf{Pr}(\cdot)$ for the probability of an event in the Borel $\sigma$-algebra on $\R^n$. The
exact projection operation on a closed convex set is denoted by $\mathcal P$, i.e., for any closed
convex set $X \subseteq \R^n$ and $\bs x \in \R^n$, we have $\mathcal P(\bs x) =
\textrm{argmin}_{\bs y \in X} \|\bs x -\bs y\|$, where $\|\cdot\|$ denotes the Euclidean norm.

\section{System Model}
We consider $M$ transmitters sharing the same media to communicate to a single receiver. We model
the channel as a Gaussian multiple access channel with flat fading effects,
\begin{equation}\label{fading_model}
    Y(n) = \sum_{i=1}^M \sqrt{H_i(n)} X_i(n) + Z(n),
\end{equation}
where $X_i(n)$ and $H_i(n)$ are the transmitted waveform and the fading process of the $i$-th
transmitter, respectively, and $Z(n)$ is properly bandlimited Gaussian noise with variance $N_0$.
We assume that the fading processes of all transmitters are jointly stationary and ergodic, and the
stationary distribution of the fading process has continuous density. We assume that all the
transmitters and the receiver have instant access to channel state information. In practice, the
receiver measures the channels and feeds back the channel information to the transmitters. The
implicit assumption in this model is that the channel variations are much slower than the data
rate, so that the channel can be measured accurately at the receiver and the amount of feedback
bits is negligible compared to that of transmitting information.

\begin{definition}\label{fading_speed_def}
The \emph{temporal variation} in fading is modeled as follows:
\begin{equation}\label{fading_speed}
    |H_i(n+1) -  H_i(n) | = V^i_n, \quad \textrm{for all } n, \ i= 1, \ldots, M,
\end{equation}
where the $V^i_n$s are nonnegative random variables independent across time slots for each $i$. We
assume that for each $i$, the random variables $V^i_n$ are uniformly bounded from above by $\hat
v^i$, which we refer to as the \emph{maximum speed of fading}. Under slow fading conditions, the
distribution of $V^i_n$ is expected to be more concentrated around zero.

\end{definition}

Consider the non-fading case where the channel state vector is fixed. The capacity region of the
Gaussian multiple-access channel with no power control is described as follows \cite{Liao},
\begin{equation}\label{Cg}
    C_g(\bs P, \bs H) = \bigg\{ \bs R \in \mathbb{R}^M_+:
    \sum_{i \in S} R_i \leq  C\Big(\sum_{i \in S} H_i P_i, N_0\Big), \textrm{for all}\  S \subseteq \mathcal M = \{1,
    \ldots, M\} \bigg\},
\end{equation}
where $P_i$ and $R_i$ are the $i$-th transmitter's power and rate, respectively. $C(P,N)$ denotes
Shannon's formula for the capacity of the AWGN channel given by
\begin{equation}\label{C_AWGN}
    C(P,N) = \frac{1}{2}\log(1+\frac{P}{N}) \quad \textrm{nats}.
\end{equation}

For a multiple-access channel with fading, but fixed transmission powers $P_i$, the
\emph{throughput} capacity region is given by averaging the instantaneous capacity regions with
respect to the fading process \cite{Shamai},
\begin{equation}\label{Ca}
    C_a(\bs P) = \bigg\{ \bs R \in \mathbb{R}^M_+: \sum_{i \in S} R_i
    \leq \mathbb{E}_{\bs H} \bigg[ C\Big(\sum_{i \in S} H_i P_i, N_0\Big) \bigg], \textrm{for all} \  S
    \subseteq \mathcal M \bigg\},
\end{equation}
where $\bs H$ is a random vector with the stationary distribution of the fading process.

A power control policy $\bs \pi$ is a function that maps any given fading state $\bs h$ to the
powers allocated to the transmitters $\bs \pi(\bs h) =(\bs \pi_1(\bs h), \ldots, \bs \pi_M(\bs
h))$. Similarly, we can define the rate allocation policy, $\mathcal R$, as a function that maps
the fading state $\bs h$ to the transmission rates, $\mathcal R(\bs h)$. For any given
power-control policy $\bs \pi$, the capacity region follows from (\ref{Ca}) as
\begin{equation}\label{Cf}
    C_f(\bs \pi) = \bigg\{\bs R \in \mathbb{R}^M_+: \sum_{i \in S} R_i \leq
    \mathbb{E}_{\bs H} \bigg[ C\Big(\sum_{i \in S} H_i \bs \pi_i(\bs H),
    N_0\Big) \bigg], \textrm {for all}\  S \subseteq \mathcal M \bigg\}.
\end{equation}
Tse \emph{et al.} \cite{Tse} have shown that the throughput capacity of a multiple access fading channel is given by 
\begin{equation}\label{C_power_ctrl}
    C(\bar{\bs P}) = \bigcup_{\bs \pi \in \mathcal{G}} C_f(\bs \pi),
\end{equation}
where $\mathcal{G} = \{ \bs \pi: \mathbb{E}_{\bs H} [\bs \pi_i(\bs H)] \leq \bar{P}_i, \textrm{for
all}\ i\} $ is the set of all power control policies satisfying the average power constraint. Let
us define the notion of boundary or dominant face for any of the capacity regions defined above.
\begin{definition}\label{dom_face}
The \emph{dominant face} or \emph{boundary} of a capacity region, denoted by $\mathcal{F}(\cdot)$,
is defined as the set of all $M$-tuples in the capacity region such that no component can be
increased without decreasing others while remaining in the capacity region.
\end{definition}

\section{Rate Allocation in a Non-fading Channel}
In this section, we address the problem of finding the optimal operation rates in a non-fading
multiple-access channel. Without loss of generality, we fix the channel state vector to unity
throughout this section, and denote the capacity region by a simpler notation $C_g(\bs P)$ instead
of $C_g(\bs P, \bs 1)$, where $\bs P>0$ denotes the transmission power. Consider the following
utility maximization problem for a $M$-user channel.
\begin{eqnarray}\label{RAC}
   \textrm{maximize} &&u(\bs R)  \nonumber \\
  \textrm{subject to}&& \bs R \in C_g(\bs P),
\end{eqnarray}
where $R_i$ and $P_i$ are $i$-th user rate and power, respectively. The utility function $u(\bs R)$
is assumed to satisfy the following conditions.
\begin{assumption}\label{assumption_u} The following conditions hold:
\begin{itemize}
\item[(a)] The utility function $u:\R^M_+ \rightarrow \R$ is concave with respect to vector $\bs R$.
\item[(b)] $u(\bs R)$ is monotonically increasing with respect to $R_i$, for $i = 1, \ldots , M$.
\end{itemize}
\end{assumption}

\begin{assumption}\label{assumption_uB}There exists a scalar $B$ such that
$$\|\bs g \| \leq B, \quad \textrm{for all}\ \bs g \in \partial u(\bs R) \hbox{ and all } \bs R,$$
where $\partial u(\bs R)$ denotes the subdifferential of $u$ at $\bs R$, i.e., the set of all
subgradients \footnote{The vector $\bs g$ is a subgradient of a concave function $f:D \rightarrow
\mathbb{R}$ at $\bs x_0$, if and only if $f(\bs x) - f(\bs x_0) \leq \bs g'(\bs x-\bs x_0)$ for all
$\bs x \in D$.} of $u$ at $\bs R$.
\end{assumption}

Note that Assumption \ref{assumption_uB} is standard in the analysis of subgradient methods for
non-differentiable optimization problems \cite{convexbook}. The maximization problem in (\ref{RAC})
is a convex program and the optimal solution can be obtained by several optimization methods such
as the gradient projection method. The gradient
    projection method with exact projection is typically used for problems where the projection operation is simple,
    i.e., for problems with simple constraint sets such as
    the non-negative orthant or a simplex. However, the constraint set in (\ref{RAC}) is defined by
    exponentially many constraints, making the projection problem computationally intractable.
    To alleviate this problem, we use an approximate projection, which is  obtained by successively projecting on violated constraints.

\begin{definition}\label{approx_proj_def}
    Let $X = \{\bs x \in \mathbb{R}^n| A\bs x \leq \bs b\}$, where $A$ has non-negative entries. Let $\bs y \in
    \mathbb{R}^n$ violate the constraint $\bs a_i' \bs x \leq b_i$, for $i\in\{i_1, \ldots, i_l\}$. The approximate
    projection of $\bs y$ on $X$, denoted by $\tilde{\mathcal P}$, is given by
    $$ \tilde{\mathcal P}(\bs y) = \mathcal P_{i_1}(\ldots (\mathcal P_{i_{l-1}}(\mathcal P_{i_l}(\bs y)))),$$
    where $\mathcal P_{i_k}$ denotes the exact projection on the hyperplane $\{\bs x \in \mathbb{R}^n| \bs a_{i_k}' \bs x =
    b_{i_k}\}$.
\end{definition}

\begin{figure}
  \centering
  \includegraphics[width=.3\textwidth]{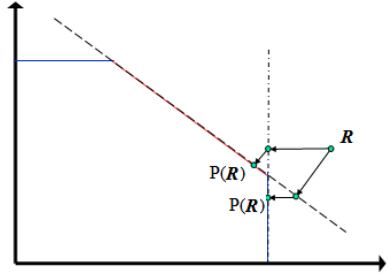}\\
  \caption{Approximate projection of $\bs R$ on a two-user MAC capacity region}\label{projection_fig}
\end{figure}

An example of approximate projection on a two-user multiple-access capacity region is illustrated
in Figure \ref{projection_fig}. As shown in the figure, the result of approximate projection is not
necessarily unique. In the following, when we write $\tilde{\mathcal P}$, it refers to an
approximate projection for an arbitrary order of projections on the violated hyperplanes. Although
the approximate projection is not unique, it is pseudo-nonexpansive as claimed in the following
Lemma.

\begin{lemma}\label{approx_proj_prop}
The approximate projection $\tilde{\mathcal P}$ given by Definition \ref{approx_proj_def} has the
following properties:
\begin{itemize}
\item[(i)] For any $\bs y \in \mathbb{R}^n$, $\tilde{\mathcal P}(\bs y)$ is feasible with respect to set $X$, i.e., $\tilde{\mathcal P}(\bs y)\in X$.
\item[(ii)] $\tilde{\mathcal P}$ is pseudo-nonexpansive, i.e.,
\begin{equation}\label{nonexpan}
    \|\tilde{\mathcal P}(\bs y) - \bs{\tl y} \| \leq \|\bs y - \bs{ \tl y}\|,\quad  \textrm{for all} \
\bs {\tl y} \in X.
\end{equation}

\end{itemize}
\end{lemma}
\begin{proof}
For part (i), it is straightforward to see that $\mathcal P_i(\bs y)$ is given by (cf. \cite{nlp}
Sec.\ 2.1.1)
$$\mathcal P_i(\bs y) = \bs y - \frac{\bs a_i'\bs y - b_i}{\|\bs a_i\|}\bs a_i.$$
Since $\bs a_i$ has only non-negative entries, all components of $\bs y$ are decreased after
projection and hence, the constraint $i$ will not be violated in the subsequent projections. This
shows that given an infeasible vector $\bs y\in \mathbb{R}^n$, the approximate projection operation
given in Definition \ref{approx_proj_def} yields a feasible vector with respect to set $X$.

Part (ii) can be verified by using the nonexpansiveness property of projection on a closed convex
set (See Proposition 2.1.3 of \cite{nlp}) for $l$ times. Since $\bs {\tl y} $ is a fixed point of
$\mathcal P_i$ for all $i$, we have
        \begin{eqnarray}
           \|\tilde{\mathcal P}(\bs y) - \bs {\tl y}\| &=&  \|\mathcal P_{i_1}(\ldots (\mathcal P_{i_l}(\bs y))) - \mathcal P_{i_1}(\ldots (\mathcal P_{i_l}(\bs {\tl y})))\| \nonumber \\
           &\leq& \|\mathcal P_{i_2}(\ldots (\mathcal P_{i_l}(\bs y))) - \mathcal P_{i_{2}}(\ldots (\mathcal P_{i_l}(\bs {\tl y})))\| \nonumber \\
           & \vdots & \nonumber \\
           &\leq& \|\bs y-\bs {\tl y}\|.
        \end{eqnarray}

\end{proof}

\begin{figure}
  \centering
  \includegraphics[width=.45\textwidth]{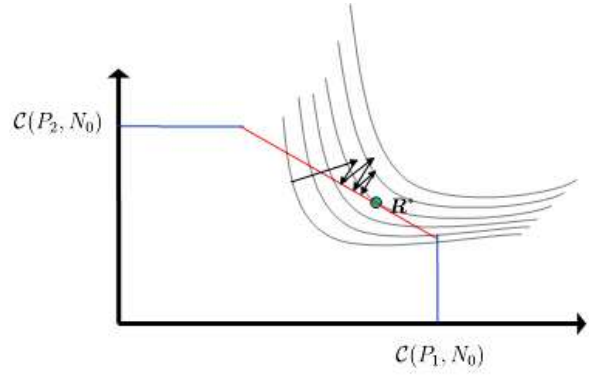}\\
  \caption{Gradient projection method with approximate projection on a two-user MAC region}\label{iterations_fig}
\end{figure}

    Here, we present the gradient projection method with approximate projection to solve the problem in (\ref{RAC}). The $k$-th iteration of the gradient projection method with approximate projection is given by
    \begin{equation}\label{iteraion}
        \bs R^{k+1} = \tilde{\mathcal P}(\bs R^k + \alpha ^k \bs g^k), \quad \bs g^k \in \partial
        u(\bs R^k),
    \end{equation}
    where $\bs g^k$ is a subgradient of $u$ at $\bs R^k$, and $\alpha ^k$ denotes the stepsize.
    Figure \ref{iterations_fig} demonstrates gradient projection iterations for a two-user multiple
    access channel. The following theorem provides a sufficient condition which can be used to
    establish convergence of (\ref{iteraion}) to the optimal solution.

    \begin{theorem}\label{convergence-thm}
        Let Assumptions \ref{assumption_u} and \ref{assumption_uB} hold, and $\bs R^*$ be an optimal solution of problem (\ref{RAC}). Also, let the sequence $\{\bs R^k\}$ be
        generated by the iteration in (\ref{iteraion}). If the stepsize $\alpha^k$ satisfies
        \begin{equation}\label{stepsize}
            0 < \alpha^k < \frac{2\left(u(\bs R^*) - u(\bs R^k)\right)}{\|\bs g^k\|^2},
        \end{equation}
then
        \begin{equation}\label{contraction}
            \|\bs R^{k+1} - \bs R^*\| < \|\bs R^k - \bs R^*\|.
        \end{equation}
    \end{theorem}

    \begin{proof}
        We have
        \begin{eqnarray}
            \|\bs R^k + \alpha ^k \bs g^k - \bs R^*\|^2 = \|\bs R^k  - \bs R^*\|^2  + 2 \alpha^k (\bs R^k - \bs R^*)' \bs g^k  + (\alpha^k)^2 \| \bs g^k\|^2. \nonumber
        \end{eqnarray}

        By concavity of $u(\cdot)$, we have
        \begin{equation}\label{conv2}
            (\bs R^* - \bs R^k)' \bs g^k \geq u(\bs R^*) - u(\bs R^k).
        \end{equation}

        Hence,
        \begin{eqnarray}
            \|\bs R^k + \alpha ^k \bs g^k - \bs R^*\|^2 \leq \|\bs R^k  - \bs R^*\|^2 - \alpha^k \left[ 2\left(u(\bs R^*) - u(\bs R^k)\right)- (\alpha^k) \| \bs g^k\|^2\right]. \nonumber
        \end{eqnarray}

    If the stepsize satisfies (\ref{stepsize}), the above relation yields the following
        $$ \|\bs R^k + \alpha ^k \bs g^k - \bs R^*\| < \|\bs R^k  - \bs R^*\|. $$

        Now by applying pseudo-nonexpansiveness of the approximate projection we have
        \begin{eqnarray}\label{conv3}
            \|\bs R^{k+1} - \bs R^*\| = \| \tilde{\mathcal P}(\bs R^k + \alpha ^k \bs g^k) - \bs R^* \| \leq \|\bs R^k + \alpha ^k \bs g^k - \bs R^*\| < \|\bs R^k  - \bs R^*\|. \nonumber
        \end{eqnarray}
    \end{proof}
\begin{theorem}\label{convergence_prop}
Let Assumptions \ref{assumption_u} and \ref{assumption_uB} hold. Also, let the sequence $\{\bs
R^k\}$ be
        generated by the iteration in (\ref{iteraion}). If the stepsize $\alpha^k$ satisfies
        (\ref{stepsize}),
then $\{\bs R^k\}$ converges to an optimal solution $\bs R^*$.
\end{theorem}

\begin{proof}
See Proposition 8.2.7 of \cite{convexbook}.
\end{proof}

The convergence analysis for this method can be extended for different stepsize selection rules.
For instance, we can employ diminishing stepsize, i.e., $$ \alpha^k \rightarrow 0, \qquad \sum_{k =
0}^{\infty} \alpha^k = \infty,$$ or more complicated dynamic stepsize selection rules such as the
\textit{path-based incremental target level} algorithm proposed by Br\"{a}nnlund \cite{brann} which
guarantees convergence to the optimal solution \cite{convexbook}, and has better convergence rate
compared to the diminishing stepsize rule.

\subsection{Complexity of the Projection Problem}
    Even though the approximate projection is simply obtained by successive projection on the
    violated constraints, it requires to find the violated constraints among exponentially many
    constraints describing the constraint set. In this part, we exploit the special structure
    of the capacity region so that each gradient projection step in (\ref{iteraion}) can be performed in
    polynomial time in $M$.

\begin{definition}\label{submodularity}
Let $f:2^\mathcal M \rightarrow \mathbb R$ be a function defined over all subsets of $\mathcal M$.
The function $f$ is \emph{submodular}  if
    \begin{equation}\label{submodular_def}
        f(S \cup T) + f(S \cap T) \leq f(S) + f(T), \quad \textrm{for all}\  S,T \in 2^\mathcal{M}.
    \end{equation}
\end{definition}

\begin{lemma}\label{submodular_lemma}
Define $f_C(S): 2^\mathcal M \rightarrow \mathbb R$ as follows:
\begin{equation}\label{fC}
f_C(S) = C\big(\sum_{i \in S} P_i, N_0\big), \quad \foral S \subseteq \mathcal M.
\end{equation}

If $P_i > 0$ for all $i \in \mathcal M$, then $f_C(S)$ is submodular. Moreover, the inequality
$(\ref{submodular_def})$ holds with equality if and only if $S \subseteq T$, or $T \subseteq S$.

\end{lemma}

\begin{proof}
The proof is simply by plugging the definition of $f_C(\cdot)$ in inequality
(\ref{submodular_def}). In particular,
\begin{eqnarray}
  f_C(S)+f_C(T) - f(S \cup T) -f(S \cap T) &=& \frac{1}{2} \log \bigg[\frac{(N_0+\sum_{i \in S} P_i) (N_0+\sum_{i \in T} P_i)}{(N_0+\sum_{i \in S \cap T} P_i) (N_0+\sum_{i \in S \cup T} P_i)} \bigg]  \nonumber \\
 &=& \frac{1}{2} \log \bigg[1+\frac{\sum_{(i,j) \in (S\setminus T)\times (T \setminus S)} P_i P_j}{(N_0+\sum_{i \in S \cap T} P_i) (N_0+\sum_{i \in S \cup T} P_i)} \bigg] \nonumber \\
 &\geq& 0.
\end{eqnarray}
 Since $P_i >0$, the above inequality holds with equality if and only if $S\setminus T = \emptyset$, or $T\setminus S =
 \emptyset$. This condition is equivalent to either $S$ or $T$ contains the other.
\end{proof}

\begin{theorem}\label{SFM_prop}
For any $\bs{\bar{R}} \in \mathbb{R}_+^M$, finding the most violated capacity constraint in
(\ref{Cg}) can be written as a \emph{submodular function minimization} (SFM) problem, that is
unconstrained minimization of a submodular function over all $S \subseteq \mathcal M$.
\end{theorem}
\begin{proof}
We can rewrite the capacity constraints of $C_g(\bs P)$ as
    \begin{equation}\label{capacity_region2}
         f_C(S) - \sum_{i \in S} R_i \geq 0, \quad \foral S
        \subseteq \mathcal{M}.
    \end{equation}

Thus, the most violated constraint at $\bs{\bar{R}}$ corresponds to
    \begin{eqnarray}\label{SFM}
        S^* = \rm arg \!\min_{S \in 2^\mathcal{M}} &&  f_C(S) -  \sum_{i \in S} \overline{R}_i. \nonumber
    \end{eqnarray}
By Lemma \ref{submodular_lemma} $f_C$ is a submodular function. Since summation of a submodular and
a linear function is also submodular, the problem above is of the form of submodular function
minimization.
\end{proof}

    It was first shown by Gr\"{o}tschel \emph{et al.} \cite{Grotschel}  that an SFM problem can be
    solved in polynomial time. The are several fully combinatorial strongly polynomial algorithms
    in the literature. The best known algorithm for SFM proposed by Orlin \cite{orlin} has running
    time $O(M^6)$. Note that approximate
    projection does not require any specific order for successive projections. Hence, finding the
    most violated constraint is not necessary for approximate projection. In view of this fact, a
    more efficient algorithm based on rate-splitting is presented in Appendix
    \ref{appendix_Rate_splitting}, to find a violated constraint. It is shown in Theorem
    \ref{ratesplit_complexity} that the rate-splitting-based algorithm runs in $O(M^2 \log M)$
    time, where $M$ is the number of users.

    Although a violated constraint can be obtained in polynomial time, it does not guarantee that
    the approximate projection can be performed in polynomial time. Because it is possible to have
    exponentially many constraints violated at some point and hence the total running time of the
    projection would be exponential in $M$. However, we show that for a small enough stepsize in the gradient
    projection iteration (\ref{iteraion}), no more than $M$ constraints can be violated at each
    iteration. Let us first define the notions of expansion and distance for a polyhedra.

        \begin{definition}\label{expansion_def}
        Let $Q$ be a polyhedron described by a set of linear inequalities, i.e.,
        \begin{equation}\label{polyhedron}
            Q = \left\{\bs x \in \mathbb{R}^n: A \bs x \leq \bs b \right\}.
        \end{equation}
        Define the \emph{expansion} of $Q$ by $\delta$, denoted by $\mathcal{E}_\delta(Q)$, as the polyhedron
        obtained by relaxing all the constraints in (\ref{polyhedron}), i.e., $ \mathcal{E}_\delta(Q) = \left\{\bs x \in \mathbb{R}^n: A \bs x \leq \bs b + \delta\mathbf{1}
            \right\},$
        where $\mathbf{1}$ is the vector of all ones.
        \end{definition}

        \begin{definition}\label{Hausdorff_def}
        Let $X$ and $Y$ be two polyhedra described by a set of linear constraints. Let
        $\mathcal{E}_d(X)$ be an expansion of $X$ by $d$ as defined in Definition \ref{expansion_def}. The distance
        $d_H(X,Y)$ between $X$ and $Y$ is defined as the minimum scalar $d$ such that $X \subseteq \mathcal{E}_d(Y)$ and $ Y \subseteq
        \mathcal{E}_d(X)$.
        \end{definition}

    \begin{lemma}\label{expansion_thm}
        Let $f_C$ be as defined in (\ref{fC}). There exists a positive scalar $\delta$ satisfying
            \begin{eqnarray}\label{exp_lemma_hyp}
                \delta \leq \frac{1}{2} (f_C(S)+ f_C(T) - f_C(S \cap T) - f_C(S \cup T)),   \qquad \foral S,T \in 2^\mathcal{M}, \quad   S   \cap T \neq
                S,T,
            \end{eqnarray}
    such that any point in the relaxed capacity region of an $M$-user multiple-access channel,
        $\mathcal{E}_{\delta}(C_g(\bs P))$, violates no more than $M$ constraints of $C_g(\bs P)$ defined in (\ref{Cg}).
    \end{lemma}

    \begin{proof}
        Existence of a positive scalar $\delta$ satisfying (\ref{exp_lemma_hyp}) follows directly from
        Lemma \ref{submodular_lemma}, using the fact that neither $S$ nor $T$ contains the other one.

        Suppose for some $\bs R \in \mathcal{E}_{\delta}(C_g(\bs P))$, there are $M+1$ violated constraints of
        $C_g(\bs P)$. Since it is not possible to have $M+1$
        non-empty nested sets in $2^\mathcal{M}$, there are at least two violated constraints corresponding to some
        sets $S,T \in 2^ \mathcal M$ where $S \cap T \neq S,T$, and
        \begin{eqnarray}
          -\sum_{i \in S} R_i &<& -f_C(S), \\
          -\sum_{i \in T} R_i &<& -f_C(T).
        \end{eqnarray}

        Since $\bs R$ is feasible in the relaxed region,
        \begin{eqnarray}\label{cap_const}
          \sum_{i \in S \cap T} R_i &\leq& f_C(S \cap T) + \delta, \\
          \sum_{i \in S \cup T} R_i &\leq& f_C(S \cup T) + \delta.
        \end{eqnarray}

            Note that if $S \cap T = \emptyset$, (\ref{cap_const}) reduces to $0 \leq \delta$, which is
            a valid inequality.

            By summing the above inequalities we conclude
            \begin{equation}\label{contrad_lemma}
                \delta > \frac{1}{2} (f_C(S)+ f_C(T) - f_C(S \cap T) - f_C(S \cup T)),
            \end{equation}
            which is a contradiction.
    \end{proof}

    \begin{theorem}\label{convtime}
Let Assumptions \ref{assumption_u} and \ref{assumption_uB} hold. Let $P_1 \leq P_2 \leq \ldots \leq
P_M$ be the transmission powers.

        If the stepsize $\alpha^k $ in the $k$-th iteration (\ref{iteraion}) satisfies
            \begin{eqnarray}\label{step_lemma_hyp}
                \alpha^k &\leq& \frac{1}{4 B \sqrt{M}} \log \bigg[1+\frac{P_1 P_2}{(N_0+\sum_{i=3}^M P_i) (N_0+\sum_{i=1}^M P_i)}
            \bigg],
            \end{eqnarray}
        then at most $M$ constraints of the capacity region $C_g(\bs P)$ can be violated at each iteration step.
    \end{theorem}
    \begin{proof} We first show that inequality in (\ref{exp_lemma_hyp}) holds for the
following choice
        of $\delta$:
        \begin{eqnarray}\label{min_delta}
            && \delta = \frac{1}{4} \log \bigg[1+\frac{P_1 P_2}{(N_0+\sum_{i=3}^M P_i) (N_0+\sum_{i=1}^M P_i)}
            \bigg].
        \end{eqnarray}
        In order to verify this, rewrite the right hand side of (\ref{exp_lemma_hyp}) as
        \begin{eqnarray}
          && \frac{1}{4} \log \bigg[\frac{(N_0+\sum_{i \in S} P_i) (N_0+\sum_{i \in T} P_i)}{(N_0+\sum_{i \in S \cap T} P_i) (N_0+\sum_{i \in S \cup T} P_i)} \bigg]  \nonumber \\
           && = \frac{1}{4} \log \bigg[1+\frac{\sum_{(i,j) \in (S\setminus T)\times (T \setminus S)} P_i P_j}{(N_0+\sum_{i \in S \cap T} P_i) (N_0+\sum_{i \in S \cup T} P_i)} \bigg] \nonumber \\
           && \geq \frac{1}{4} \log \bigg[1+\frac{P_1 P_2}{(N_0+\sum_{i \in S \cap T} P_i) (N_0+\sum_{i \in S \cup T} P_i)} \bigg] \nonumber \\
           && \geq \frac{1}{4} \log \bigg[1+\frac{P_1 P_2}{(N_0+\sum_{i \in S \cap T} P_i) (N_0+\sum_{i=1}^M P_i)} \bigg] \nonumber \\
           && \geq \frac{1}{4} \log \bigg[1+\frac{P_1 P_2}{(N_0+\sum_{i=3}^M P_i) (N_0+\sum_{i=1}^M P_i)} \bigg]. \nonumber
        \end{eqnarray}

        The inequalities can be justified by using the monotonicity of the logarithm function and the fact
        that $(S\setminus T)\times (T \setminus S)$ is non-empty because $S \cap T \neq S,T$.

        Now, let $\bs R^k$ be feasible in the capacity region, $C_g(\bs P)$. For every $S \subseteq \mathcal M$, we have
        \begin{eqnarray}\label{stp_thm_1}
          \sum_{i \in S} (R^k_i + \alpha^k g^k_i) &=& \sum_{i \in S} R^{k}_i + \alpha^k \|g^{k}\|  \sum_{i \in S} \frac{g^{k}_i}{\|g^{k}\| } \nonumber \\
          &\leq& f(S) + \frac{\delta}{B \sqrt{M}} B \sum_{i \in S} \frac{g^{k}_i}{\|g^{k}\| }
          \nonumber \\
          &\leq& f(S) + \delta,
        \end{eqnarray}
        where the first inequality follows from Assumption \ref{assumption_u}(b), Assumption
        \ref{assumption_uB}, and Eq. (\ref{step_lemma_hyp}). The second inequality holds because
        for any unit vector  $\bs d \in \mathbb R^M$, it is true that
        \begin{equation}\label{stp_thm_2}
            \sum_{i \in S} d_i \leq \sum_{i \in S} |d_i| \leq \sqrt{M}.
        \end{equation}

        Thus, if $\alpha^k$ satisfies (\ref{step_lemma_hyp}) then $(\bs R^k + \alpha^k \bs g^k) \in
        \mathcal{E}_{\delta}(C_g(\bs P))$, for some $\delta$ for which (\ref{exp_lemma_hyp}) holds.
        Therefore, by Lemma \ref{expansion_thm} the number of violated
        constraints does not exceed $M$.
        \end{proof}

In view of the fact that a violated constraint can be identified in $O(M^2\log M)$ time (see the
Algorithm in Appendix \ref{appendix_Rate_splitting}), Theorem \ref{convtime} implies that, for
small enough stepsize, the approximate projection can be implemented in $O(M^3\log M)$ time.

In section \ref{nopower_sec}, we will develop algorithms that use the gradient projection method
for dynamic rate allocation in a time varying channel.

\section{Dynamic Rate and Power Allocation in Fading Channel with Known Channel
Statistics}\label{power_sec}

        In this section, we assume that the channel statistics are known. Our goal is to find feasible rate and power allocation policies denoted
        by $\mathcal{R}^*$ and $\bs \pi^*$, respectively, such that $\mathcal{R}^*(\bs H) \in C_g\big(\bs \pi^*(\bs H),\bs
        H\big)$, and $\bs \pi^* \in \mathcal G$. Moreover,
    \begin{equation}\label{RAC_pctrl}
       \mathbb{E}_{\boldsymbol{H}} [\mathcal{R}^*(\bs H)] = \bs R^* \in \textrm{argmax} \quad u(\bs R), \qquad \textrm{subject to} \quad \bs R \in  C(\bar{\bs
       P}),
      \end{equation}
        where $u(\cdot)$ is a given utility function and is assumed to be differentiable and satisfy Assumption
        \ref{assumption_u}.

        For the case of a linear utility function, i.e., $u(\boldsymbol R) = \boldsymbol \mu '
    \boldsymbol R $ for some $\boldsymbol \mu \in \mathbb{R} _{+}^M $, Tse \emph{et al.} \cite{Tse}
    have  shown that the optimal rate and power allocation policies are given by the optimal solution to a
    linear program, i.e.,
        \begin{equation}\label{LP_RAC_pctrl}
            \left(\mathcal R^*(\bs h), \bs \pi^*(\bs h)\right) = \textrm{arg}\max_{\boldsymbol r , \boldsymbol p} \left( \boldsymbol \mu ' \boldsymbol r - \boldsymbol
            \lambda ' \boldsymbol p \right), \quad \textrm{subject to} \quad \boldsymbol r \in
            C_g( \boldsymbol p,\boldsymbol h),
        \end{equation}
        where $\boldsymbol h$ is the channel state realization, and $\boldsymbol \lambda \in
        \mathbb{R} _{+}^M$ is a Lagrange multiplier satisfying the average power constraint, i.e.,
        $\bs \lambda$ is the unique solution of the following equations

        \begin{equation}\label{lambda_mu}
            \int_0^{\infty}\!\! \frac{1}{h} \int_{\frac{2
            \lambda_i (N_0+z) }{\mu_i}}^{\infty} \prod_{k \neq i} F_k \left( \frac{2 \lambda_k h
            (N_0 + z)}{2 \lambda_i (N_0+z) + (\mu_k - \mu_i) h} \right) f_i(h) \mathrm{d}h
            \mathrm{d}z = \bar{P}_i,
        \end{equation}
        where $F_k$ and $f_k$ are, respectively, the cumulative distribution function (CDF) and
        the probability density function (PDF) of the stationary distribution of the channel state
        process for transmitter $k$.

        Exploiting the polymatroid structure of the capacity region, problem (\ref{LP_RAC_pctrl})
        can be solved by a simple greedy algorithm (see Lemma 3.2 of \cite{Tse}). It is also shown
        in \cite{Tse} that, for positive $\boldsymbol \mu$, the optimal solution, $\boldsymbol
        R^*$, to the problem in (\ref{RAC_pctrl}) is \emph{uniquely} obtained. Given the
        distribution of channel state process, denoted by $F_k$ and $f_k$, we have
        \begin{equation}\label{R_mu}
            R_i^*(\boldsymbol \mu) = \int_0^{\infty}\!\!\!\! \frac{1}{2(N_0+z)} \int_{\frac{2
            \lambda_i (N_0+z) }{\mu_i}}^{\infty} \prod_{k \neq i} F_k \left( \frac{2 \lambda_k h
            (N_0 + z)}{2 \lambda_i (N_0+z) + (\mu_k - \mu_i) h} \right) f_i(h) \mathrm{d}h
            \mathrm{d}z.
        \end{equation}

        The uniqueness of $\bs R^*$ follows from the fact that the stationary distribution of the
        channel state process has a continuous density \cite{Tse}. It is worth mentioning that
        (\ref{R_mu}) parametrically describes the \emph{boundary} of the capacity region which is
        precisely defined in Definition \ref{dom_face}. Thus, there is a one-to-one correspondence
        between the boundary of $C(\bs{\bar{P}})$ and the positive vectors $\bs \mu$ with unit
        norm.

        Now consider a general concave utility function satisfying Assumption \ref{assumption_u}.
        It is straightforward to show that $\bs R^*$, the optimal solution to (\ref{RAC_pctrl}), is unique.
        Moreover, by Assumption \ref{assumption_u}(b) it lies on the boundary of the throughput region.
        Now suppose that $\bs R^*$ is given by some genie. We can choose $\bs \mu^* = \nabla u(\bs{R}^*)$ and $\tilde u(\bs R) =
        (\bs \mu^*)'\bs R$, as a replacement for the nonlinear utility. By checking the optimality
        conditions, it can be seen that $\bs R^*$ is also the optimal solution of the problem in
        (\ref{RAC_pctrl}), i.e.,
        \begin{equation}\label{R_mu_corresp}
            \bs R^* = \textrm{argmax} \quad (\bs\mu ^*)' \bs R \quad \textrm{subject to} \quad \bs R \in
            C(\bar{\bs P}).
        \end{equation}

        Thus, we can employ the greedy
        rate and power allocation policies in (\ref{LP_RAC_pctrl}) for the linear utility function
        $\tilde u(\cdot)$, and achieve the optimal average rate for the nonlinear utility function
        $u(\cdot)$.  Therefore, the problem of optimal resource allocation reduces to computing the vector $\bs
        R^*$. Note that the throughput capacity region is not characterized by a finite set of
        constraints, so standard optimization methods such as gradient projection or interior-point methods are not applicable in
        this case. However, the closed-form solution to maximization of a linear function on the
        throughput region is given by (\ref{R_mu}). This naturally leads us to the conditional
        gradient method \cite{nlp} to compute $\bs R^*$. The $k$-th iteration of the method is given by
        \begin{equation}\label{frank-wolfe}
            \bs R^{k+1} = \bs R^k + \alpha^k(\bs{\bar{R}}^k - \bs R^k),
        \end{equation}
        where $\alpha^k $ is the stepsize and $\bs{\bar{R}}^k$ is obtained as
        \begin{equation}\label{frank-wolfe2}
            \bs{\bar{R}}^k \in \textrm{arg}\!\!\!\!\max_{\bs{R} \in C(\bs{\bar{P}})} \left( \nabla u(\bs{R}^k)'(\bs
            R - \bs R^k)\right),
        \end{equation}
        where $\nabla u(\bs{R}^k)$ denotes the gradient vector of $u(\cdot)$ at $\bs{R}^k$. Since
        the utility function is monotonically increasing by Assumption \ref{assumption_u}(b), the
        gradient vector is always positive and, hence, the unique optimal solution to the above
        sub-problem is obtained by (\ref{R_mu}), in which $\bs \mu$ is replaced by $\nabla
        u(\bs{R}^k)$. By concavity of the utility function and convexity of the capacity region,
        the iteration (\ref{frank-wolfe}) will converge to the optimal solution of
        (\ref{RAC_pctrl}) for appropriate stepsize selection rules such as the Armijo rule or limited
        maximization rule (cf. \cite{nlp} pp. 220-222).

        Note that our goal is to determine rate and power allocation policies. Finding $\bs R^*$
        allows us to determine such policies by the greedy policy in  (\ref{LP_RAC_pctrl}) for $\bs \mu
        ^* = \nabla u(\bs{R}^*)$. It is worth mentioning that all the computations for obtaining $\bs
        R^*$ are performed once in the setup of the communication session. Here, the convergence rate of the conditional
        gradient method is generally not of critical importance.

\section{Dynamic Rate Allocation without Knowledge of Channel Statistics}\label{nopower_sec}
        In this part we assume that the channel statistics are not known and that the
        transmission powers are fixed to $\bs P$. In practice, this scenario occurs when the
        transmission power may be limited owing to environmental limitations such as human presence,
        or limitations of the hardware.

        The capacity region of the fading multiple access channel for this scenario is a polyhedron
        given by (\ref{Ca}). Similarly to the previous case, the goal is to find an optimal rate
        allocation policy $\mathcal R^*(\cdot)$ with respect to a given utility function, which we formally define next.

\begin{definition}\label{optimal_policy}
[Optimal Policy] The optimal rate allocation policy denoted by $\mathcal{R}^*(\cdot)$ is a mapping
that satisfies $\mathcal{R}^*(\bs H) \in C_g\big(\bs P,\bs H\big)$ for all $\bs H$, such that
\begin{eqnarray}\label{RAC_np}
\mathbb{E}_{\boldsymbol{H}} [\mathcal{R}^*(\bs H)] = \bs R^* \in& \textrm{argmax}& \quad u(\bs R)
\nonumber \\
&\textrm{subject to}& \quad \bs R \in  C_a({\bs P}).
\end{eqnarray}
\end{definition}

        It is worth noting that the approach used to find the optimal resource allocation
        policies for the case with known channel statistics does not apply to this scenario, because $C_g(\bs
        P, \bs h)$ is a polyhedron and hence, unlike in Section \ref{power_sec} the uniqueness of the optimal solution, $\bs R^*$ for any positive vector
        $\bs \mu$ does not hold anymore.

        Here we present a greedy rate allocation policy and compare its performance with the
        unknown optimal policy. The performance of a particular rate allocation policy is defined
        as the utility function evaluated at the average rate achieved by that policy.

\begin{definition}\label{greedy_policy}
[Greedy Policy] A \emph{greedy} rate allocation policy, denoted by $\mathcal{\bar{R}}$, is given by
\begin{eqnarray}\label{RAC_greedy}
\mathcal{\bar{R}}(\bs H) = & \textrm{argmax}& \quad u(\bs R)
\nonumber \\
&\textrm{subject to}& \quad \bs R \in  C_g({\bs P}, \bs H)
\end{eqnarray}
i.e., for each channel state, the greedy policy chooses the rate vector that maximizes the utility
function over the corresponding capacity region.
\end{definition}

The utility function $u(\bs R)$ is assumed to satisfy the following conditions.

\begin{assumption}\label{assumption_u2} For every $\delta > 0$, let $\mathcal N_\delta =\Big\{ \bs H: d_H(C_g(\bs P,\bs H), C_a(\bs P)) \leq \delta\Big\}$. The following conditions hold:
\begin{itemize}
\item[(a)] There exists a scalar $B(\delta)$ such that for all $\bs H\in \mathcal N_\delta$,
$$              |u(\bs{R}_1) - u(\bs{R}_2)| \leq B(\delta) \|\bs{R}_1 - \bs{R}_2\|, \qquad \foral \ \bs{R}_i, \|\bs R_i\| \geq D_\delta, i=1,2,$$
where
 \begin{equation}\label{diameter}
    D_\delta = \inf_{\bs H \in \mathcal N_\delta} \sup_{\bs R \in C_g(\bs P,\bs H)} \|\bs R\|.
 \end{equation}

\item[(b)] There exists a scalar $A(\delta)$ such that for all $\bs H\in \mathcal N_\delta$,
$$              |u(\bar{\mathcal{R}}(\bs H)) - u(\bs{R})| \geq A(\delta) \|\bar{\mathcal{R}}(\bs H) - \bs{R}\|^2, \quad \foral \  \bs{R} \in C_g(\bs P, \bs H). $$

\end{itemize}
\end{assumption}

Assumption \ref{assumption_u2}(a) is a weakened version of Assumption \ref{assumption_uB}, which
imposes a bound on subgradients of the utility function. This assumption only requires bound on the
subgradient in a neighborhood of the optimal solution and away from the origin, which is satisfied
by a larger class of functions. Assumption \ref{assumption_u2}(b) is a strong concavity type
assumption. In fact, strong concavity of the utility implies Assumption \ref{assumption_u2}(b), but
it is not necessary. The scalar $A(\delta)$ becomes small as the utility tends to have a linear
structure with level sets tangent to the dominant face of the capacity region. Assumption
\ref{assumption_u2} holds for a large class of utility functions including the well known
$\alpha$-fair functions given by
\begin{equation}\label{a_fair}
    f_{\alpha}(x) = \left \{ \begin{array}{ll}
                   \frac{x^{1-\alpha}}{1-\alpha}, & \textrm{$\alpha \neq 1$}\\
                   \log(x), & \textrm{$\alpha = 1,$}
                   \end{array} \right.
\end{equation}
which do not satisfy Assumption \ref{assumption_uB}.

Note that the greedy policy is not necessarily optimal for general concave utility functions.
Consider the following relations
        \begin{eqnarray}\label{jensen}
          \mathbb{E}_{\bs H}\big[u\big(\mathcal{R}^*(\bs H)\big)\big] &\leq& \mathbb{E}_{\bs H}\big[u\big(\bar{\mathcal{R}}(\bs H)\big)\big] \nonumber \\
          &\leq& u\big(\mathbb{E}_{\bs H}\big[\bar{\mathcal{R}}(\bs H)\big]\big) \nonumber \\
          &\leq& u\big(\mathbb{E}_{\bs H}\big[\mathcal{R}^*(\bs H)\big]\big),
        \end{eqnarray}
        where the first and third inequality follow from the feasibility of the optimal and the
        greedy policy for any channel state, and the second inequality follows from Jensen's
        inequality by concavity of the utility function.

        In the case of a linear utility function we have $u\big(\mathbb{E}_{\bs H}\big[\mathcal{R}^*(\bs
        H)\big]\big) =  \mathbb{E}_{\bs H}\big[u\big(\mathcal{R}^*(\bs H)\big)\big]$, so equality holds throughout in (\ref{jensen}) and
        $\bar{\mathcal{R}}(\cdot)$ is indeed the optimal rate allocation policy. For
        nonlinear utility functions, the greedy policy can be strictly suboptimal.

        However, the greedy policy is not arbitrarily worse than the optimal one. In view of (\ref{jensen}), we can bound the
        performance difference, $u(\bs{R}^*) - u\big(\mathbb{E}_{\bs H}\big[\bar{\mathcal{R}}(\bs H)\big]\big)$, by
        bounding $\Big|u\big(\mathbb{E}_{\bs H}\big[\mathcal{R}^*(\bs H)\big]\big) - u\big(\mathbb{E}_{\bs H}\big[\bar{\mathcal{R}}(\bs H)\big]\big)\Big|$ or
        $\Big|u\big(\mathbb{E}_{\bs H}\big[\mathcal{R}^*(\bs H)\big]\big) -  \mathbb{E}_{\bs H}\big[u\big(\mathcal{R}^*(\bs
        H)\big)\big]\Big|$ from above. We show that the first bound goes to zero as the channel variations
        become small and the second bound vanishes as the utility function tends to have a more
        linear structure.

        Before stating the main theorems, let us introduce some useful lemmas. The first lemma
        asserts that the optimal and greedy policies assign rates on the dominant face of the
        capacity region.

        \begin{lemma}\label{dom_face_lemma}
            Let $u(\cdot)$ satisfy Assumption \ref{assumption_u}(b). Also, let
           ${\mathcal{R}^*}(\cdot)$ and $\bar{\mathcal{R}}(\cdot)$ be optimal and greedy rate allocation
           policies as in Definitions \ref{optimal_policy} and \ref{greedy_policy}, respectively. Then,
           \begin{itemize}
               \item[(a)] $\bar{\mathcal{R}}(\bs H) \in \mathcal F\big(C_g(\bs P, \bs H)\big), \quad \foral \bs H.$
               \item[(b)] $\textbf{\textrm{Pr}} \big\{\bs H: {\mathcal{R}^*}(\bs H) \in \mathcal F\big(C_g(\bs P, \bs H)\big) \big\} =1.$
           \end{itemize}
            where $\mathcal F(\cdot)$ denotes the dominant face of a capacity region (cf. Definition
            \ref{dom_face}).
        \end{lemma}

        \begin{proof}
            Part (a) is direct consequence of Assumption \ref{assumption_u}(b) and Definition
            \ref{dom_face}. If the optimal solution to the utility maximization
            problem is not on the dominant face, there exists a user $i$ such that we can increase
            its rate and keep all other user's rates fixed while staying in the capacity region. Thus,
            we are able to increase the utility by Assumption \ref{assumption_u}(b), which leads to a contradiction.

            For part (b), first note that with the same argument as above we have
            \begin{eqnarray}\label{dom_face_eq1}
                 \bs R^* = \mathbb{E}_{\boldsymbol{H}} [\mathcal{R}^*(\bs H)] &\in& \mathcal F\big(C_a(\bs
                 P)\big).
            \end{eqnarray}

            From Definition \ref{dom_face} and the definition of throughput capacity region in (\ref{Ca}),
            we have
            \begin{equation}\label{dom_face_eq2}
                 \mathbb{E}_{\boldsymbol{H}} \Big[\sum _{i=1}^M \mathcal{R}_i^*(\bs
                 H)\Big]  =  \mathbb{E}_{\boldsymbol{H}} \Big[ C\Big(\sum _{i=1}^M H_i P_i, N_0\Big)\Big].
            \end{equation}

            Thus, $\sum _{i=1}^M \mathcal{R}_i^*(\bs H) = C\big(\sum _{i=1}^M H_i P_i, N_0\big)$, with
            probability one, because $C\big(\sum _{i=1}^M H_i P_i, N_0\big) - \sum _{i=1}^M \mathcal{R}_i^*(\bs H) \geq 0$, for all $\bs H$. Therefore, by definition of MAC capacity region in (\ref{Cg}) we conclude ${\mathcal{R}^*}(\bs H) \in \mathcal F\big(C_g(\bs P, \bs
            H)\big)$, with probability one.
        \end{proof}

         The following lemma extends Chebyshev's inequality for capacity regions. It states that, with high probability,
        the time varying capacity region does not deviate much from its mean.

        \begin{lemma} \label{region_chebyshev}
            Let $\bs H$ be a random vector with the stationary distribution of the channel state
            process, mean $\bs{\bar{H}}$ and covariance matrix $K$. Then
            \begin{equation}\label{capacity_cheby}
                \textbf{\textrm{Pr}} \Big\{ d_H \left(C_g(\bs{P},\bs{H}), C_a(\bs{P}) \right) > \delta \Big\}
                \leq \frac{\sigma_H^2}{\delta^2},
            \end{equation}
            where $\sigma_H^2$ is defined as
              \begin{equation}\label{sigma_H}
                           \sigma_H^2 \triangleq \frac{1}{4}\sum_{S \subseteq \{1,\ldots, M\}} \bs{\Gamma}_S' K \bs \Gamma_S \left(1+ \left[(1+\bs{\Gamma}'_S \bs{\bar{H}})(\sqrt{2 \log(1+\bs{\Gamma}'_S \bs{\bar{H}})} -
                             \frac{\sqrt{\bs \Gamma_S' K \bs \Gamma_S}}{2})\right]^2\right),
              \end{equation}
                where
                  \begin{equation}\label{P_indicator}
                   {(\bs \Gamma_S)}_i = \left\{ \begin{array}{ll}
                   \frac{P_i}{N_0}, & \textrm{$i \in S$}\\
                   0, & \textrm{otherwise.}
                   \end{array} \right.
                 \end{equation}
        \end{lemma}
\begin{proof}
See Appendix \ref{chebyshev_pf}.
\end{proof}

        The system parameter $\sigma_H^2$ in Lemma \ref{region_chebyshev} is proportional to
        channel variations, and we expect it to vanish for very small channel variations. The
        following lemma ensures that the distance between the optimal solutions of the utility
        maximization problem over two regions is small, provided that the regions are close to each
        other.
\begin{lemma}\label{opt_dist}
Let the utility function, $u: \R^M \rightarrow \R$, satisfy Assumptions \ref{assumption_u} and
\ref{assumption_u2}. Also, let $\bs R_1^*$ and $\bs R_2^*$ be the optimal solution of maximizing
the utility over $C_a(\bs P)$ and $C_g(\bs P, \bs H)$, respectively. If
$$d_H\big(C_g(\bs P, \bs H), C_a(\bs P)\big) \leq \delta,$$
then we have
\begin{equation}\label{opt_dist_result}
    \|\bs R_1^* - \bs R_2^*\| \leq
    {\delta}^{\frac{1}{2}}\left[{\delta}^{\frac{1}{2}}+\Big(\frac{B(\delta)}{A(\delta)}\Big)^{\frac{1}{2}}\right].
\end{equation}

\end{lemma}

\begin{proof}
See Appendix \ref{opt_dist_pf}.
\end{proof}

    The following theorem combines the results of the above two lemmas to obtain a bound on the performance
    difference of the greedy and the optimal policy.
\begin{theorem}\label{Bound2}
    Let $u:\R^M \rightarrow \R_+$ satisfy Assumptions \ref{assumption_u} and \ref{assumption_u2}. Also, let ${\mathcal{R}^*}(\cdot)$ and $\bar{\mathcal{R}}(\cdot)$ be optimal and greedy rate allocation policies as
    in Definitions \ref{optimal_policy} and \ref{greedy_policy}, respectively.  Then for every $\delta \in
    [\sigma_H^2,\infty)$,
    \begin{equation}\label{Bound2_eps}
        u(\bs{R}^*) - u\big(\mathbb{E}_{\bs H}\big[\bar{\mathcal{R}}(\bs H)\big]\big) \leq
        \frac{\sigma_H^2}{\delta^2} u(\bs{R}^*) +
          \Big(1-\frac{\sigma_H^2}{\delta^2}\Big)B(\delta)\Big[\delta^{\frac{1}{2}}+\Big(\frac{B(\delta)}{A(\delta)}\Big)^{\frac{1}{2}}\Big]\delta^{\frac{1}{2}},
    \end{equation}
    where $\bs R^* =\mathbb{E}_{\boldsymbol{H}} [\mathcal{R}^*(\bs H)]$, and $A(\delta)$ and
    $B(\delta)$ are positive scalars defined in Assumption \ref{assumption_u2}.
\end{theorem}

        \begin{proof}
            Pick any $\delta \in [\sigma_H^2,\infty)$. Define the event, $\mathcal{V}$ as
            $$\mathcal{V} = \bigg\{ \bs H: d_H\big(C_g(\bs P, \bs H), C_a(\bs P)\big)  \leq \delta\bigg\}.$$

              By Lemma \ref{region_chebyshev}, the probability of this event is at least $1-\frac{\sigma_H^2}{\delta^2}$. Using Jensen's inequality as in (\ref{jensen}) we can bound the left-hand side of
            (\ref{Bound2_eps}) as follows
            \begin{eqnarray}\label{bound2_ch1}
              u(\bs{R}^*) - u\big(\mathbb{E}_{\bs H}[\bar{\mathcal{R}}(\bs H)]\big) &\leq& u(\bs{R}^*) - \mathbb{E}_{\bs H}\big[u(\bar{\mathcal{R}}(\bs H))\big] \nonumber \\
               &=& u(\bs{R}^*) - (1-\frac{\sigma_H^2}{\delta^2})\mathbb{E}_{\bs H}\Big[u(\bar{\mathcal{R}}(\bs H))\Big| \mathcal{V}\Big] \nonumber \\
               & & - \mathbf{Pr}(\mathcal{V}^c)\mathbb{E}_{\bs H}\Big[u(\bar{\mathcal{R}}(\bs H))\Big| \mathcal{V}^c\Big] \nonumber \\
               &\leq& \frac{\sigma_H^2}{\delta^2} u(\bs{R}^*) + (1-\frac{\sigma_H^2}{\delta^2})\bigg( u(\bs{R}^*) - \mathbb{E}_{\bs H}\big[u(\bar{\mathcal{R}}(\bs
               H))| \mathcal{V}\big] \bigg) \nonumber \\
               &\leq& \frac{\sigma_H^2}{\delta^2} u(\bs{R}^*) + (1-\frac{\sigma_H^2}{\delta^2})\bigg| \mathbb{E}_{\bs H}\Big[u(\bs{R}^*) - u(\bar{\mathcal{R}}(\bs
               H))\big| \mathcal{V}\Big] \bigg| \nonumber \\
               &\leq& \frac{\sigma_H^2}{\delta^2} u(\bs{R}^*) + (1-\frac{\sigma_H^2}{\delta^2})\mathbb{E}_{\bs H}\Big[ |u(\bs{R}^*) -u(\bar{\mathcal{R}}(\bs H)) | \Big| \mathcal{V}
               \Big].
            \end{eqnarray}
            In the above relations, the first inequality follows from the fact that
            $\mathbf{Pr}(\mathcal{V}) \geq 1- \frac{\sigma_H^2}{\delta^2}$, and the second inequality holds because of
            the non-negativity of $u(\bs R)$.

            On the other hand, by incorporating Lemma \ref{dom_face_lemma} in Assumption \ref{assumption_u2}(a) we have
                $$ |u(\bs{R}^*) -u(\bar{\mathcal{R}}(\bs H)) | \leq B(\delta) \|\bar{\mathcal{R}}(\bs H) - \bs{R}^*\|. $$

            Now by Assumption \ref{assumption_u2} we can employ Lemma \ref{opt_dist} to
            conclude the following from the above relation:
            \begin{eqnarray}\label{bound2_ch2}
              |u(\bs{R}^*) -u(\bar{\mathcal{R}}(\bs H)) | &\leq&  B(\delta)\Big(\delta^{\frac{1}{2}}+\Big(\frac{B(\delta)}{A(\delta)}\Big)^{\frac{1}{2}}\Big)\delta^{\frac{1}{2}},\quad \foral \bs H, d_H\big(C_g(\bs P, \bs H), C_a(\bs P)\big) \leq \delta, \nonumber
            \end{eqnarray}
            which implies
            \begin{eqnarray}
              \mathbb{E}_{\bs H}\Big[ \big|u(\bs{R}^*) -u(\bar{\mathcal{R}}(\bs H)) \big| \Big|
              \mathcal{V}\Big] &\leq&
              B(\delta)\Big(\delta^{\frac{1}{2}}+\Big(\frac{B(\delta)}{A(\delta)}\Big)^{\frac{1}{2}}\Big)\delta^{\frac{1}{2}}.
            \end{eqnarray}

            The desired result follows immediately from substituting (\ref{bound2_ch2}) in
            (\ref{bound2_ch1}).

        \end{proof}

\begin{figure}
  \centering
  \includegraphics[width=.5\textwidth]{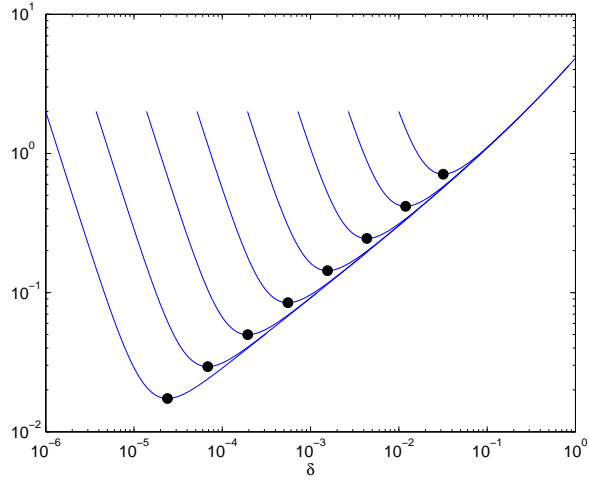}\\
  \caption{Parametric upper bound on performance difference between greedy and optimal policies as
  in right hand side of (\ref{Bound2_eps}) for different channel variations,
  $\sigma_H$, as a function of $\delta$}\label{Bound2_fig}
\end{figure}

        Theorem \ref{Bound2} provides a bound parameterized by $\delta$. For very small channel
        variations, $\sigma_H$ becomes small. Therefore, the parameter $\delta$ can be picked small
        enough  such that the bound in (\ref{Bound2_eps}) tends to zero. Figure \ref{Bound2_fig}
        illustrates the behavior of right hand side of Eq. (\ref{Bound2_eps}) as a function of
        $\delta$ for different values of $\sigma_H$. For each value of $\sigma_H$, the upper bound
        is minimized for a specific choice of $\delta$, which is illustrated by a dot in Figure
        \ref{Bound2_fig}. As demonstrated in the figure, for smaller channel variations, a smaller
        gap is achieved and the parameter $\delta$ that minimizes the bound decreases.

        The next theorem provides another bound  demonstrating the impact of the structure of the
        utility function on the performance of the greedy policy.

        \begin{theorem}\label{Bound1}
            Let Assumption \ref{assumption_u} hold for the twice differentiable function $u:\R^M
    \rightarrow \R_+$. Also, let ${\mathcal{R}^*}(\cdot)$ and $\bar{\mathcal{R}}(\cdot)$ be the optimal
    and the greedy rate allocation policies, defined in Definitions \ref{optimal_policy} and
    \ref{greedy_policy}, respectively. Then for every $\epsilon \in (0,1]$,
            \begin{equation}\label{Bound1_eps}
                u(\bs{R}^*) - u\big(\mathbb{E}_{\bs H}\big[\bar{\mathcal{R}}(\bs H)\big]\big) \leq \epsilon
                u(\bs{R}^*) +
                \frac{1}{2}(1-\epsilon) r(\epsilon)^2 \Omega,
            \end{equation}
            where $\bs R^* =\mathbb{E}_{\boldsymbol{H}} [\mathcal{R}^*(\bs H)]$, and $\Omega$ satisfies the following
            \begin{equation}\label{bound2_hyp1}
                \lambda_{\max}\big(- \nabla^2 u(\bs \xi)\big) \leq \Omega, \qquad \textrm{for all}\  \bs \xi,\ \|\bs \xi - \bs
                R^*\| \leq r(\epsilon),
            \end{equation}
            in which $\nabla^2$ denotes the Hessian of $u$, and $r(\epsilon)$ is given by
            \begin{equation}\label{bound1_hyp2}
                r(\epsilon) =  \sqrt{M}\frac{\sigma_H}{\sqrt{\epsilon}} + \left[ \sum_{i=1}^M \mathbb{E}_{\bs H}\left[\frac{1}{2} \log \left(\frac{(1+H_i P_i)(1+\sum_{j \neq i} H_j P_j)}{1+\sum_{j=1}^M H_j P_j}\right)\right]^2
                \right]^\frac{1}{2}.
            \end{equation}
        \end{theorem}

           \begin{proof}
                Similarly to the proof of Theorem \ref{Bound2}, for any $\epsilon \in (0,1]$ define the event
                $\mathcal{V}$ as
                 \begin{equation}\label{event_V}
                                 \mathcal{V} = \bigg\{ \bs H:d_H(C_g(\bs P, \bs H), C_a(\bs P)) \leq
                                 \frac{\sigma_H}{\sqrt{\epsilon}}\bigg\}.
                 \end{equation}

                By Lemma \ref{region_chebyshev}, this event has probability at least $1-\epsilon$.
                Lemma \ref{dom_face_lemma} asserts that the optimal policy  almost surely allocate rate vectors on
                the dominant face of $C_g(\bs P,\bs H)$. Therefore, for almost all $\bs H \in \mathcal{V}$, the optimal policy satisfies the following
                \begin{equation}\label{R_opt_interval}
                   \mathbb{E}_{\bs H}\bigg[\frac{1}{2} \log\Big(1+\frac{H_i P_i}{1+\sum_{j \neq i} H_j P_j}\Big)\bigg]- \frac{\sigma_H}{\sqrt{\epsilon}} \leq \mathcal{R}_i^*(\bs H) \leq \mathbb{E}_{\bs H}\bigg[\frac{1}{2} \log\Big(1+H_i
                   P_i\Big)\bigg] + \frac{\sigma_H}{\sqrt{\epsilon}}.
                \end{equation}

                Thus, for almost all $\bs H \in \mathcal V$, we have
                $$ |\mathcal{R}_i^*(\bs H) - R_i^*| \leq \frac{\sigma_H}{\sqrt{\epsilon}} + \mathbb{E}_{\bs H}\left[\frac{1}{2} \log \left(\frac{(1+H_i P_i)(1+\sum_{j \neq i} H_j P_j)}{1+\sum_{j=1}^M H_j P_j}\right)\right].$$

                Therefore,
                \begin{eqnarray}\label{Rs_r_neighbor}
                  \|{\mathcal{R}}^*(\bs H) - \bs R^*\|&\leq& \sqrt{M}\frac{\sigma_H}{\sqrt{\epsilon}} + \left[ \sum_{i=1}^M \mathbb{E}_{\bs H}\left[\frac{1}{2} \log \left(\frac{(1+H_i P_i)(1+\sum_{j \neq i} H_j P_j)}{1+\sum_{j=1}^M H_j P_j}\right)\right]^2
                \right]^\frac{1}{2} \nonumber \\
                &=& r(\epsilon), \quad \textrm{for almost all } \bs H \in \mathcal V.
                \end{eqnarray}

                Now let us write the Taylor expansion of $u(\cdot)$ at $\bs R^*$ in
                the direction of $\bs R$,
                \begin{eqnarray}
                  u(\bs R) &=& u(\bs R^*) + \nabla u(\bs R^*)'(\bs R - \bs R^*) - \frac{1}{2}(\bs R - \bs R^*)'(-\nabla^2u(\xi))(\bs R - \bs R^*) \nonumber \\
                  &\geq& u(\bs R^*) + \nabla u(\bs R^*)'(\bs R - \bs R^*) - \frac{1}{2}\|\bs R - \bs R^*\|^2 \lambda_{\max}(-\nabla^2u(\xi)) \nonumber \\
                  && \qquad \qquad \qquad \qquad  \qquad \textrm{for some} \  \xi, \ \|\xi - \bs R^*\| \leq \|\bs R -\bs
                  R^*\|.
                \end{eqnarray}

                In the above relation, let $\bs R = \mathcal{R}^*(\bs H)$ for all $\bs H \in \mathcal{V}$. The utility function is concave, so its Hessian is negative
                definite and we can combine (\ref{Rs_r_neighbor}) with the above relation to write
                \begin{eqnarray}
                   u(\mathcal{R}^*(\bs H))&\geq& u(\bs R^*) + \nabla u(\bs R^*)'( \mathcal{R}^*(\bs H)- \bs R^*)
                   - \frac{1}{2} r(\epsilon)^2 \Omega, \quad \textrm{for almost all } \bs H \in
                   \mathcal V.
                \end{eqnarray}

                Taking the expectation conditioned on $\mathcal{V}$, and using the fact that
                $\mathcal{R}^*(\bs H) \in \mathcal F\big(C_g(\bs P,\bs H)\big)$ we have the
                following
                \begin{equation}\label{U_Rs_expectatoin}
                  \mathbb{E}_{\bs H}\big[u(\mathcal{R}^*(\bs H))\big|\mathcal{V}\big] \geq u(\bs R^*) - \frac{1}{2} r(\epsilon)^2
                  \Omega.
                \end{equation}

                Hence, we conclude
              \begin{eqnarray}\label{bound1_ch1}
              u(\bs{R}^*) - u(\mathbb{E}_{\bs H}(\bar{\mathcal{R}}(\bs H))) &\leq& u(\bs{R}^*) - \mathbb{E}_{\bs H}[u(\mathcal{R}^*(\bs H))] \nonumber \\
               &\leq& u(\bs{R}^*) - (1-\epsilon)\mathbb{E}_{\bs H}\Big[u({\mathcal{R}^*}(\bs H))\Big| \mathcal{V}\Big] \nonumber \\
               & & - \mathbf{Pr}(\mathcal{V}^c)\mathbb{E}_{\bs H}\Big[u({\mathcal{R}^*}(\bs H))\Big| \mathcal{V}^c\Big] \nonumber \\
               &\leq& u(\bs{R}^*) - (1-\epsilon)\big(u(\bs R^*) - \frac{1}{2} r(\epsilon)^2 \Omega \big) \nonumber \\
               &=& \epsilon u(\bs{R}^*) + \frac{1}{2}(1-\epsilon) r(\epsilon)^2 \Omega. \nonumber
            \end{eqnarray}
            where the first inequality is verified by (\ref{jensen}), and the third inequality
            follows from non-negativity of the utility function and the inequality in (\ref{U_Rs_expectatoin}).
            \end{proof}

            Similarly to Theorem \ref{Bound2}, Theorem \ref{Bound1} provides a bound parameterized
            by $\epsilon$. As the utility function tends to have a  more linear structure, $\Omega$
            tends to zero. For instance, $\Omega$ is proportional to $\alpha$ for a weighted sum $\alpha$-fair utility function.
            Hence, we can choose $\epsilon$ small such that the right hand side of
            (\ref{Bound1_eps}) goes to zero. The behavior of this upper bound for different values
            of $\Omega$ is similar to the one plotted in Figure \ref{Bound2_fig}.

            In summary, the performance difference between the greedy and the optimal policy is
            bounded from above by the minimum of the bounds provided by Theorem \ref{Bound2} and
            Theorem \ref{Bound1}.

Even though the greedy policy can perform closely to the optimal policy, it requires solving a
nonlinear program in each time slot. For each channel state, finding even a near-optimal solution
of the problem in (\ref{RAC_greedy}) requires a large number of iterations, making the online
evaluation of the greedy policy impractical. In the following section, we introduce an alternative
rate allocation policy, which implements a single gradient projection iteration of the form
(\ref{iteraion}) per time slot.

\subsection{Approximate Rate Allocation Policy}
In this part, we assume that the channel state information is available at each time slot $n$, and
the computational resources are limited such that a single iteration of the gradient projection
method in (\ref{iteraion}) can be implemented in each time slot. In order to simplify the notation
in this part and avoid unnecessary technical details, we consider a stronger version of Assumption
\ref{assumption_u2}(b).

\begin{assumption}\label{assumption_u3}
\emph{Let $\bs R^\dag = \textrm{argmax}_{\bs R \in C_g(\bs P, \bs H)} u(\bs R)$. Then there exists
a positive scalar $A$ such that
$$|u(\bs R^\dag) - u(\bs R)| \geq A \|\bs R^\dag - \bs R\|^2, \quad \textrm{for all } \bs R \in C_g(\bs P, \bs H).$$
}\label{utility_assump3}
\end{assumption}


\begin{definition}\label{approximate_policy}
[Approximate Policy] Given some fixed integer $k \geq 1$, we define the \emph{approximate} rate
allocation policy, $\widetilde{\mathcal{R}}$, as follows:

   \begin{equation}\label{policy_worst}
    \widetilde{\mathcal{R}}\big(\bs H(n)\big) \triangleq \left\{ \begin{array}{ll}
    \bar{\mathcal{R}}\big(\bs H(0)\big), & n = 0\\
    \bs{\tilde R}_{t(n)}^\tau, & n \geq 1,
    \end{array}\right.
  \end{equation}
where
\begin{equation}\label{tau_max}
    \tau = \textrm{arg}\!\!\!\!\!\!\max_{0 \leq j < k-1} \ u(\bs{\tilde R}^j_{t(n)}), \quad {t(n)} = \left\lfloor \frac{n-1}{k} \right\rfloor,
\end{equation}
and $\bs{\tilde R}^j_{t(n)} \in \mathbb R^M$ is given by the following gradient projection
iterations:
\begin{eqnarray}\label{iteration_tv}
    \bs{\tilde R}_{t(n)}^0 &=& \tilde P_{{t(n)}}\left[ \widetilde{\mathcal{R}}\Big(\bs H\big(k t(n)\big)\Big)\right],  \nonumber \\
    \bs{\tilde R}_{t(n)}^{j+1} &=& \tilde P_{{t(n)}}\left[ \bs{\tilde R}_{t(n)}^j + \alpha^j \bs{\tilde g}_{t(n)}^j
    \right], \quad j=1,\ldots,k-1,
\end{eqnarray}
where $\bs{\tilde g}_{t(n)}^j$ is a subgradient of $u(\cdot)$ at $\bs{\tilde R}_{t(n)}^j$,
$\alpha^j$ denotes the stepsize and $\tilde P_{{t(n)}}$ is the approximate projection on
$C_g\big(\bs P, \bs H(kt(n))\big)$.
\end{definition}

For $k=1$, (\ref{iteration_tv}) reduces to taking only one gradient projection iteration at each
time slot. For $k >1$, the proposed rate allocation policy essentially allows the channel state to
change for a block of $k$ consecutive time slots, and then takes $k$ iterations of the gradient
projection method with the approximate projection. We will show below that this method tracks the
greedy policy closely. Hence, this yields an efficient method that on average requires only one
iteration step per time slot. Note that to compute the policy at time slot $n$, we are using the
channel state information at time slots $kt, k(t-1), \ldots$. Hence, in practice the channel
measurements need to be done only every $k$ time slots.

There is a tradeoff in choosing system parameter $k$, because taking only one gradient projection
step may not be sufficient to get close enough to the greedy policy's operating point. Moreover,
for large $k$ the new operating point of the greedy policy can be far from the previous one, and
$k$ iterations may be insufficient.

Before stating the main result, let us introduce some useful lemmas. In the following lemma, we
translate the model in Definition \ref{fading_speed_def} for temporal variations in channel state
into changes in the corresponding capacity regions.

\begin{lemma}\label{region_dist}
Let $\big\{[H_i(n)]_{i=1,\ldots,M}\big\}$ be the fading process that satisfies condition in
(\ref{fading_speed}). We have
\begin{equation}\label{capacity_dist}
    d_H\Big(C_g\big(\bs P, \bs H(n+1)\big),C_g\big(\bs P, \bs H(n)\big)\Big) \leq W_n,
\end{equation}
where $\{W_n\}$ are non-negative independent identically distributed random variables bounded from
above by $\hat w = \frac{1}{2} \sum_{i=1}^M \hat v^i P_i$, where $\hat v^i$ is a uniform upper
bound on the sequence of random variables $\{V_n^i\}$ and $P_i$ is the $i$-th user's transmission
power.
\end{lemma}
\begin{proof}
By Definition \ref{Hausdorff_def} we have
\begin{eqnarray}\label{region_lemma1}
    &&d_H\Big(C_g\big(\bs P, \bs H(n+1)\big),C_g\big(\bs P, \bs H(n)\big)\Big)  \nonumber \\
    &&= \max_{S \subseteq \mathcal M} \frac{1}{2} \bigg |\log\Big( 1+ \frac{\sum_{i \in S}(H_i(n+1)-H_i(n))P_i}{1+ \sum_{i \in S} H_i(n)P_i}
    \Big)\bigg|  \nonumber \\
    && \leq \max_{S \subseteq \mathcal M} \frac{\sum_{i \in S}|H_i(n+1)-H_i(n)|P_i}{2(1+ \sum_{i \in S}
    H_i(n)P_i)} \nonumber \\
    && \leq \frac{1}{2} \sum_{i=1}^M|H_i(n+1)-H_i(n)|P_i =\frac{1}{2} \sum_{i=1}^M V^i_n P_i.
\end{eqnarray}
Therefore, (\ref{capacity_dist}) is true for $W_n = \frac{1}{2} \sum_{i=1}^M V^i_n P_i$. Since the
random variables $V^i_n$ are i.i.d. and bounded above by $\hat v^i_n$, the random variables $W_n$
are i.i.d. and bounded from above by $\frac{1}{2} \sum_{i=1}^M \hat v^i P_i$.

\end{proof}

The following useful lemma by Nedi\'c and Bertsekas \cite{subgradient_rate} addresses the
convergence rate of the gradient projection method with constant stepsize.

\begin{lemma}\label{subgradient_rate}
Let rate allocation policies $\bar{\mathcal{R}}$  and $\widetilde{\mathcal{R}}$ be given by
Definition \ref{greedy_policy} and Definition \ref{approximate_policy}, respectively. Also, let
Assumptions \ref{assumption_u}, \ref{assumption_uB} and \ref{assumption_u3} hold and the stepsize
$\alpha^n$ be fixed to some positive constant $\alpha$. Then for a positive scalar $\epsilon$ we
have
\begin{equation}\label{greedy_worst_rate}
    u\Big(\widetilde{\mathcal{R}}\big(\bs H(n)\big)\Big) \geq  u\Big(\bar{\mathcal{R}}\big(\bs
    H(kt)\big)\Big) - \frac{\alpha B^2 + \epsilon}{2},
\end{equation}
if $k$ satisfies
\begin{equation}\label{k_condition}
    k \geq \bigg \lfloor \frac{\|\bs{\tilde R}_t^{0} - \bar{\mathcal{R}}\big(\bs
    H(kt)\big) \|^2}{\alpha \epsilon} \bigg \rfloor.
\end{equation}

\end{lemma}

\begin{proof}
See Proposition 2.3 of \cite{subgradient_rate}.
\end{proof}

We next state our main result, which shows that the approximate rate allocation policy given by
Definition \ref{approximate_policy} tracks the greedy policy within a neighborhood which is
quantified as a function of the maximum speed of fading, the parameters of the utility function,
and the transmission powers.

\begin{theorem}\label{tracking_worst}
Let Assumptions \ref{assumption_u}, \ref{assumption_uB} and \ref{assumption_u3} hold and the rate
allocation policies $\bar{\mathcal{R}}$ and $\widetilde{\mathcal{R}}$ be given by Definition
\ref{greedy_policy} and Definition \ref{approximate_policy}, respectively. Choose the system
parameters $k$ and $\alpha$ for the approximate policy in Definition \ref{approximate_policy} as
$$k = \bigg\lfloor(\frac{2B}{A w'})^{\frac{2}{3}}\bigg\rfloor, \qquad \alpha = \bigg(\frac{16A}{B^2}\bigg)^\frac{1}{3} {w'}^\frac{2}{3},$$
where $w' = {\hat w}^{\frac{1}{2}}\big(\hat w^{\frac{1}{2}}+(\frac{B}{A})^{\frac{1}{2}}\big)$,
$\hat w$ is the upper bound on $W_n$ as defined in Lemma \ref{region_dist}, $A$ and $B$ are
constants given in Assumptions \ref{assumption_u3} and \ref{assumption_uB}. Then, we have
\begin{equation}\label{greedy_worst_dist}
    \|\widetilde{\mathcal{R}}\big(\bs H(n)\big) - \bar{\mathcal{R}}\big(\bs H(n)\big)\| \leq
    2\theta = 2\Big(\frac{2B}{A}\Big)^{\frac{2}{3}} w'^{\frac{1}{3}}.
\end{equation}

\end{theorem}

\begin{proof}
First, we show that
\begin{equation}\label{tracking1}
    \|\widetilde{\mathcal{R}}\big(\bs H(n)\big) - \bar{\mathcal{R}}\big(\bs H(kt)\big)\| \leq
    \theta = \Big(\frac{2B}{A}\Big)^{\frac{2}{3}} w'^{\frac{1}{3}},
\end{equation}
where $t = \lfloor \frac{n-1}{k} \rfloor$. The proof is by induction on $t$. For $t=0$ the claim is
trivially true. Now suppose that (\ref{tracking1}) is true for some positive $t$. Hence, it also
holds for $n = k(t+1)$ by induction hypothesis, i.e.,
\begin{equation}\label{tracking2}
   \|\bs{\tilde R}_{t+1}^{0} - \bar{\mathcal{R}}\big(\bs H(kt)\big) \| \leq \theta.
\end{equation}

On the other hand, by Lemma \ref{region_dist} implies that for every $n$,
$$    d_H\Big(C_g\big(\bs P, \bs H(n+1)\big),C_g\big(\bs P, \bs H(n)\big)\Big) \leq \hat w.$$

Thus, by Lemma \ref{opt_dist} and the triangle inequality we have
\begin{equation}\label{tracking3}
   \|\bar{\mathcal{R}}\big(\bs H(k(t+1))\big) - \bar{\mathcal{R}}\big(\bs H(kt)\big) \| \leq kw'
   \leq \theta.
\end{equation}

Therefore, by another triangle inequality we conclude from (\ref{tracking2}) and (\ref{tracking3})
that
\begin{equation}\label{tracking4}
       \|\bs{\tilde R}_{t+1}^{0} - \bar{\mathcal{R}}\big(\bs H(k(t+1))\big) \| \leq 2\theta.
\end{equation}

After plugging the corresponding values of $\alpha$ and $\theta$, it is straightforward to show
that (\ref{k_condition}) holds for $\epsilon = \alpha B^2$. Thus, we can apply Lemma
\ref{subgradient_rate} to show
\begin{equation}\label{tracking5}
    \bigg|u\Big(\widetilde{\mathcal{R}}\big(\bs H(n)\big)\Big) - u\Big(\bar{\mathcal{R}}\big(\bs
    H(k(t+1))\big)\Big)\bigg| \leq \alpha B^2.
\end{equation}

By Assumption \ref{assumption_u3} we can write
\begin{equation}\label{tracking6}
    \|\widetilde{\mathcal{R}}\big(\bs H(n)\big) - \bar{\mathcal{R}}\big(\bs H(k(t+1))\big)\| \leq
    \Big(\frac{\alpha B^2}{A}\Big)^\frac{1}{2} = \theta.
\end{equation}

Therefore, the proof of (\ref{tracking1}) is complete by induction.

Again by applying Lemma \ref{opt_dist} and Lemma \ref{region_dist} we have
\begin{equation}\label{tracking7}
   \|\bar{\mathcal{R}}\big(\bs H(n)\big) - \bar{\mathcal{R}}\big(\bs H(kt)\big) \| \leq kw'
   \leq \theta,
\end{equation}
and the desired result directly follows from (\ref{tracking1}) and (\ref{tracking7}) by the
triangle inequality.
\end{proof}


Theorem \ref{tracking_worst} provides a bound on the size of the tracking neighborhood as a
function of the maximum speed of fading, denoted by $\hat w$, which may be too conservative. It is
of interest to provide a rate allocation policy and a bound on the size of its tracking
neighborhood as a function of the average speed of fading. The next section addresses this issue.

\subsection{Improved Approximate Rate Allocation Policy} In this section, we design an efficient
rate allocation policy that tracks the greedy policy within a neighborhood characterized by the
average speed of fading which is typically much smaller than the maximum speed of fading. We
consider policies which can implement one gradient projection iteration per time slot.

Unlike the approximate policy given by (\ref{policy_worst}) which uses the channel state
information once in every $k$ time slots, we present an algorithm which uses the channel state
information in all time slots. Roughly speaking, this method takes a fixed number of gradient
projection iterations only after the change in the channel state has reached a certain threshold.

\begin{definition}\label{improved_approximate_policy}
[Improved Approximate Policy] Let $\{W_n\}$ be the  sequence of non-negative random variables as
defined in Lemma \ref{region_dist}, and $\gamma$ be a positive constant. Define the sequence
$\{T_i\}$ as
\begin{eqnarray}\label{T_def}
  T_0 &=& 0, \nonumber \\
  T_{i+1} &=& \min\bigg\{ t \mid \sum_{n = T_i}^{t-1} W_n \geq \gamma \bigg\}.
\end{eqnarray}

Define the \emph{improved approximate} rate allocation policy, $\widehat{\mathcal{R}}$, with
parameters $\gamma$ and $k$, as follows:
   \begin{equation}\label{policy_avg}
    \widehat{\mathcal{R}}\big(\bs H(n)\big) \triangleq \left\{ \begin{array}{ll}
    \bar{\mathcal{R}}\big(\bs H(0)\big), & n = 0\\
    \bs{\hat R}_{t(n)}^\tau, & n \geq 1,
    \end{array}\right.
  \end{equation}
where
\begin{eqnarray}\label{tau_avg}
    t(n) &=& \max\{i \mid T_i < n\}, \\
    \tau &=& \textrm{arg}\!\!\!\!\!\max_{0 \leq j < k-1} \ u\Big(\bs{\hat R}^j_{t(n)}\Big),
\end{eqnarray}
and $\bs{\hat R}^j_{t(n)} \in \mathbb R^M$ is given by the following gradient projection iterations
\begin{eqnarray}\label{iteration_tv_avg}
    \bs{\hat R}_{t(n)}^0 &=& \tilde P_{{t(n)}}\left[ \widehat{\mathcal{R}}\big(\bs H(T_{t(n)})\big)\right],  \nonumber \\
    \bs{\hat R}_{t(n)}^{j+1} &=& \tilde P_{{t(n)}}\left[ \bs{\hat R}_{t(n)}^j + \alpha^j \bs{\hat g}_{t(n)}^j
    \right], \quad j=1,\ldots,k-1,
\end{eqnarray}
where $\bs{\hat g}_{t(n)}^j$ is a subgradient of $u(\cdot)$ at $\bs{\hat R}_{t(n)}^j$, $\alpha^j$
denotes the stepsize and $\tilde P_{{t(n)}}$ is the approximate projection on $C_g(\bs P, \bs
H(T_{t(n)}))$.
\end{definition}

\begin{figure}
  \centering
  \includegraphics[width=.5\textwidth]{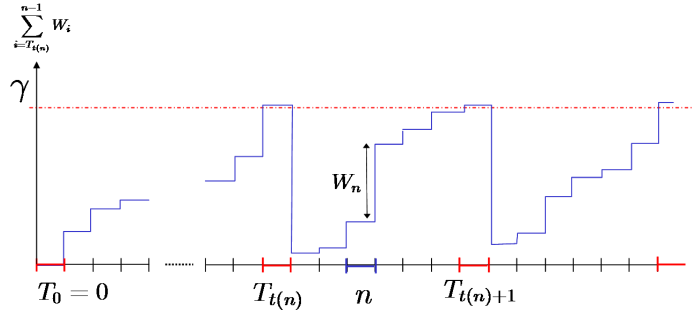}\\
  \caption{The improved approximate policy takes $k$ gradient projection iterations at time
  $T_{t(n)}$, which is the time that the random walk generated by the random variables $W_n$ reach the threshold
  $\gamma$.}\label{imp_approx_fig}
\end{figure}

Figure \ref{imp_approx_fig} depicts a particular realization of the random walk generated by $W_n$,
and the operation of the improved approximate policy.

\begin{theorem}\label{iteration_queue}
Let ${t(n)}$ be as defined in (\ref{tau_avg}), and let $\bar w= \mathbb{E}[W_n]$. If $k =
\frac{\gamma}{\bar w}$, then we have
\begin{equation}\label{almost_sure_nk}
    \lim_{n \rightarrow \infty} \frac{n}{t(n)k} = 1, \quad \textrm{with probability } 1.
\end{equation}

\end{theorem}

\begin{proof}
The sequence $\{T_i\}$ is obtained as the random walk generated by the $W_n$ crosses the threshold
level $\gamma$. Since the random variables $W_n$ are positive, we can think of the threshold
crossing as a renewal process, denoted by $N(\cdot)$, with inter-arrivals $W_n$.

We can rewrite the limit as follows
\begin{equation}\label{queue1}
     \lim_{n \rightarrow \infty} \frac{n-N\big({t(n)}\gamma\big) + N\big({t(n)}\gamma\big)}{t(n)k} =  \lim_{n \rightarrow \infty}
     \frac{n-N\big({t(n)}\gamma\big)}{t(n)k} + \bar w \frac{N\big({t(n)}\gamma\big)}{t(n)\gamma}.
\end{equation}

Since the random walk will hit the threshold with probability 1, the first term goes to zero with
probability 1. Also, by Strong law for renewal processes the second terms goes to 1 with
probability 1 (see \cite{dsp}, p.60).
\end{proof}

Theorem \ref{iteration_queue} essentially guarantees that the number of gradient projection
iterations is the same as the number of channel measurements in the long run with probability 1.

\begin{theorem}\label{tracking_avg}
Let Assumptions \ref{assumption_u}, \ref{assumption_uB} and \ref{assumption_u3} hold and the rate
allocation policies $\bar{\mathcal{R}}$ and $\widehat{\mathcal{R}}$ be given by Definition
\ref{greedy_policy} and Definition \ref{improved_approximate_policy}, respectively. Also, let $k =
\lfloor \frac{\gamma}{\bar w}\rfloor$, and fix the stepsize to $\alpha = \frac{A\gamma^2}{B^2}$ in
(\ref{iteration_tv_avg}), where $\gamma = c(\frac{B}{A})^\frac{3}{4} \bar w^\frac{1}{4}$, and $c
\geq 1$ is a constant satisfying the following equation
\begin{equation}\label{c_equation}
    \frac{(c^2-1)^8}{2^8 c^4} = \hat w.
\end{equation}
Then
\begin{equation}\label{tracking_avg_dist}
    \|\widehat{\mathcal{R}}\big(\bs H(n)\big) - \bar{\mathcal{R}}\big(\bs H(n)\big)\| \leq
    2\gamma + \Big(\frac{\gamma B}{A}\Big)^\frac{1}{2}.
\end{equation}

\end{theorem}

\begin{proof}
We follow the line of proof of Theorem \ref{tracking_worst}. First, by induction on $t$ we show
that
\begin{equation}\label{tracking_avg1}
 \|\widehat{\mathcal{R}}\big(\bs H(n)\big) - \bar{\mathcal{R}}\big(\bs H(T_t)\big)\| \leq
    \gamma,
\end{equation}
where $t$ is defined in (\ref{tau_avg}). The base is trivial. Similar to (\ref{tracking2}), by
induction hypothesis we have
\begin{equation}\label{tracking_avg2}
   \|\bs{\hat R}_{t+1}^{0} - \bar{\mathcal{R}}\big(\bs H(T_t)\big) \| \leq \gamma.
\end{equation}
By definition of $T_i$ in (\ref{T_def}) we can write
\begin{equation}\label{tracking_avg3}
    d_H\bigg(C_g\Big(\bs P, \bs H(T_{t+1})\Big),C_g\Big(\bs P, \bs H(T_t)\Big)\bigg) \leq \gamma.
\end{equation}
Thus, by Lemma \ref{opt_dist}, we have
\begin{equation}\label{tracking_avg4}
     \|\bar{\mathcal{R}}\big(\bs H(T_{t+1})\big) - \bar{\mathcal{R}}\big(\bs H(T_t)\big) \| \leq
     \gamma^\frac{1}{2}\Big(\gamma^\frac{1}{2} + \big(\frac{B}{A}\big)^\frac{1}{2}\Big).
\end{equation}
Therefore, by combining (\ref{tracking_avg2}) and (\ref{tracking_avg4}) by triangle inequality we
obtain
\begin{equation}\label{tracking_avg5}
   \|\bs{\hat R}_{t+1}^{0} - \bar{\mathcal{R}}\big(\bs H(T_{t+1})\big) \| \leq 2\gamma + \Big(\frac{\gamma
   B}{A}\Big)^\frac{1}{2}.
\end{equation}

Using the fact that $\bar w \leq \hat w = \frac{(c^2-1)^8}{2^8 c^4}$, after a few steps of
straightforward manipulations we can show that
\begin{equation}\label{tracking_avg6}
    \|\bs{\hat R}_{t+1}^{0} - \bar{\mathcal{R}}\big(\bs H(T_{t+1})\big) \|^2 \leq \Big(2\gamma + \big(\frac{\gamma B}{A}\big)^\frac{1}{2}\Big)^2 \leq c^4 \frac{\gamma
    B}{A}.
\end{equation}

Now by plugging the values of $\alpha$ and $\gamma$ in terms of system parameters in
(\ref{k_condition}), we can verify that
\begin{equation}\label{tracking_avg7}
    k = \left\lfloor \frac{\gamma}{\bar w}\right\rfloor =  \bigg\lfloor \frac{c^4\frac{\gamma B}{A}}{A\frac{\gamma^2}{B^2} A\gamma^2}
    \bigg\rfloor \geq \bigg\lfloor \frac{\|\bs{\hat R}_{t+1}^{0} - \bar{\mathcal{R}}\big(\bs H(T_{t+1})\big) \|^2}{\alpha \epsilon}
    \bigg\rfloor.
\end{equation}

Hence, we can apply Lemma \ref{subgradient_rate} for $\epsilon = A\gamma^2$, and conclude
\begin{equation}\label{tracking_avg8}
    \bigg|u\Big(\widehat{\mathcal{R}}\big(\bs H(n)\big)\Big) - u\Big(\bar{\mathcal{R}}\big(\bs
    H(T_{t+1})\big)\Big)\bigg| \leq \alpha B^2.
\end{equation}

By exploiting Assumption \ref{assumption_u3} we have
\begin{equation}\label{tracking_avg9}
    \|\widehat{\mathcal{R}}\big(\bs H(n)\big) - \bar{\mathcal{R}}\big(\bs H(T_{t+1})\big)\| \leq
    \Big(\frac{\alpha B^2}{A}\Big)^\frac{1}{2} = \gamma.
\end{equation}

Therefore, the proof of (\ref{tracking_avg1}) is complete by induction. Similarly to
(\ref{tracking_avg4}) we have
\begin{equation}\label{tracking_avg10}
     \|\bar{\mathcal{R}}\big(\bs H(n)\big) - \bar{\mathcal{R}}\big(\bs H(T_t)\big) \| \leq
     \gamma^\frac{1}{2}\Big(\gamma^\frac{1}{2} + (\frac{B}{A})^\frac{1}{2}\Big),
\end{equation}
and (\ref{tracking_avg_dist}) follows immediately from (\ref{tracking_avg1}) and
(\ref{tracking_avg10}) by invoking triangle inequality.
\end{proof}

Theorem \ref{iteration_queue} and Theorem \ref{tracking_avg} guarantee that the presented rate
allocation policy tracks the greedy policy within a small neighborhood while only one gradient
projection iteration is computed per time slot, with probability 1. The neighborhood is
characterized in terms of the average behavior of temporal channel variations and vanishes as the
fading speed decreases.

\section{Simulation Results and Discussion}
In this section, we provide simulation results to complement our analytical results and make a
comparison with other fair resource allocation algorithms. We focus on the case with no power
control or knowledge of channel statistics. We also make reasonable assumption that the channel
state processes are generated by independent identical finite state Markov chains. We consider
weighted $\alpha$-fair function as the utility function, i.e.,
\begin{equation}\label{a_fair_util}
    u(\bs R) = \sum_i{w_i f_\alpha(R_i)},
\end{equation}
where $f_\alpha(\cdot)$ is given by equation (\ref{a_fair}).

\begin{figure}
  \centering
  \includegraphics[width=.5\textwidth]{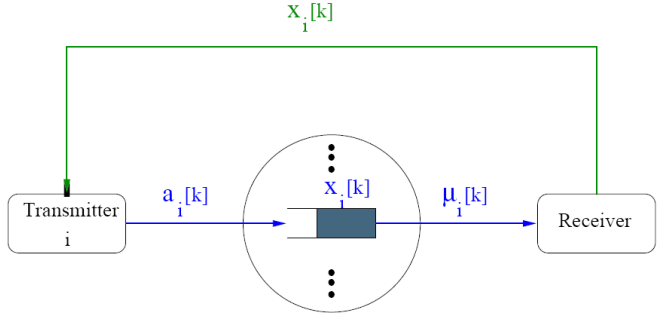}\\
  \caption{Structure of the $i$-th transmitter and the receiver for the queue-length-based policy \cite{erysri05}.}\label{queue_fig}
\end{figure}

\begin{figure}
  \centering
  \includegraphics[width=.4\textwidth]{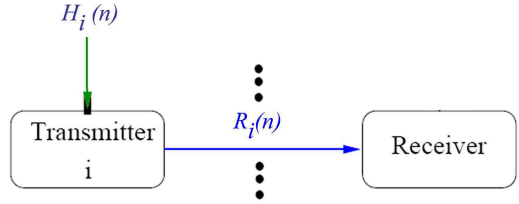}\\
  \caption{Structure of the $i$-th transmitter and the receiver for the presented policies.}\label{queue_fig}
\end{figure}

We consider two different scenarios to compare the performance of the greedy policy with the
queue-based rate allocation policy by Eryilmaz and Srikant \cite{erysri05}. This policy,
parameterized by some parameter $K$, uses queue length information to allocate the rates
arbitrarily close to the optimal policy by choosing $K$ large enough. As illustrated in Figure
\ref{queue_fig}, $x_i(n)$ denotes the queue-length of the $i$-th user. At time slot $n$, the
scheduler chooses the service rate vector $\bs \mu (n)$ based on a max-weight policy, i.e.,
\begin{eqnarray}\label{MW_scheduler}
   \bs \mu(n) = &\textrm{argmax}& \sum_{i=1}^M x_i(n)R_i
\nonumber \\
&\textrm{subject to}& \quad \bs R \in  C_g({\bs P}, \bs H(n))
\end{eqnarray}

The congestion controller proposed in \cite{erysri05} leads to a fair allocation of the rates for a
given $\alpha$-fair utility function. In particular, the data generation rate for the $i$-th
transmitter, denoted by $a_i(n)$ is a random variable satisfying the following conditions:
\begin{eqnarray}
  \mathbb E\big[a_i(n)\ | x_i(n)\big] &=& \min \bigg\{K\Big(\frac{w_i}{x_i(n)}\Big)^{\frac{1}{\alpha}}, D
  \bigg\}, \nonumber \\
    \mathbb E\big[a_i^2(n)\ | x_i(n)\big] &\leq& U < \infty, \quad \foral x_i(n),
\end{eqnarray}
where $\alpha$, $D$ and $U$ are positive constants.

\begin{figure}
  \centering
  \includegraphics[width=.5\textwidth]{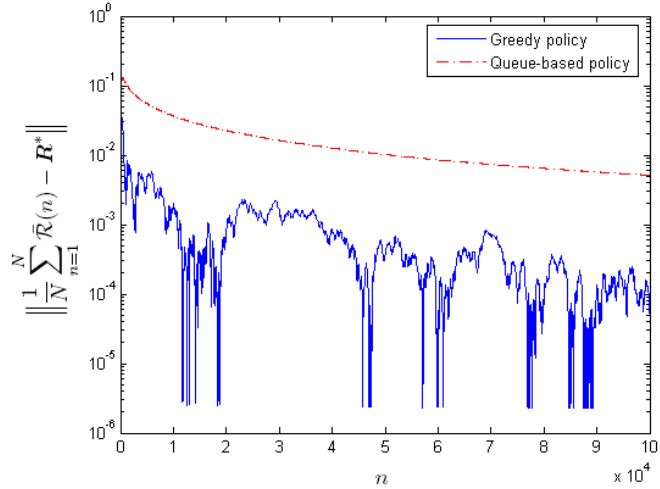}\\
  \caption{Performance comparison of greedy and queue-based policies for a communication session with limited duration, for $\frac{\sigma_H}{\bar H} = 1.22$.}\label{greedy_Q_fig}
\end{figure}

\begin{figure}
  \centering
  \includegraphics[width=.5\textwidth]{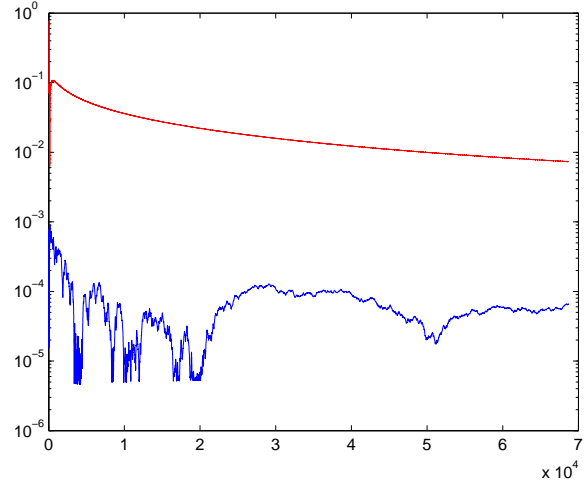}\\
  \caption{Performance comparison of greedy and queue-based policies for a communication session with limited duration, for $\frac{\sigma_H}{\bar H} = 0.13$.}\label{greedy_Q_fig2}
\end{figure}

In the first scenario, we compare the average achieved rate of the policies for a communication
session with limited duration. Figure \ref{greedy_Q_fig} depicts the distance between empirical
average  rate achieved by the greedy or the queue-length based policy, and $\bs R^*$, the maximizer
of the utility function over the throughput region. In this case, the utility function is given by
(\ref{a_fair_util}) with $\alpha = 2$ and $w_1 = 1.5w_2 = 1.5$, and the corresponding optimal
solution is $\bs R^* = (0.60, 0.49)$. As observed in Figure \ref{greedy_Q_fig}, the greedy policy
outperforms the queue-length based policy a communication session with limited duration. It is
worth noting that there is a tradeoff in choosing the parameter $K$ of the queue-length based
policy. In order to guarantee achieving close to optimal rates by queue-based policy, the parameter
$K$ should be chosen large which results in large expected queue length and lower convergence rate.
On the other hand, if $K$ takes a small value to improve the convergence rate, the achieved rate of
the queue based policy converges to a larger neighborhood of the $\bs R^*$.

As established in Theorem \ref{Bound2}, the performance of the greedy policy improves by decreasing
the channel variations. Figure \ref{greedy_Q_fig2} demonstrates the improvement in performance of
the greedy policy when $\frac{\sigma_H}{\bar{H}}$ decreases from 1.22 to 0.13. We also observe in
Figure \ref{greedy_Q_fig2} that the queue-length based policy is not sensitive to channel
variations, and its performance does not improve by decreasing the channel variations. It is worth
mentioning that the greedy policy as observed in the simulation results performs significantly
better than the bounds provided by Theorems \ref{Bound2} and \ref{Bound1}. These upper bounds
characterize the behavior of the greedy policy in terms of channel variations and structure of the
utility function, but they are not necessarily tight.

\begin{figure}
  \centering
  \includegraphics[width=.5\textwidth]{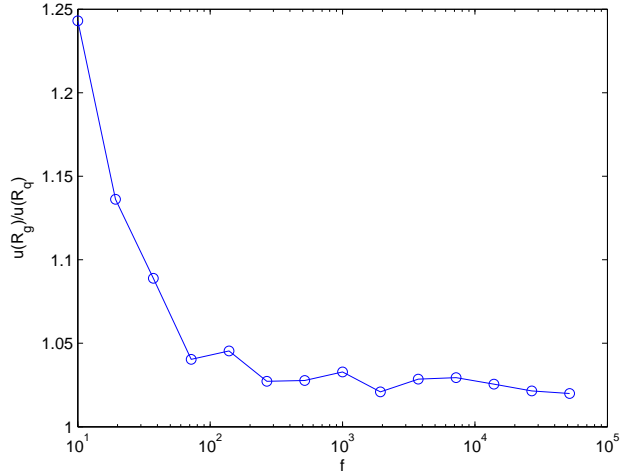}\\
  \caption{Performance comparison of greedy and queue-based policies for file upload scenario with
  respect to file size $f = f_1 = f_2$. $R_g$ and $R_q$ are expected upload rate of the greedy and
  the queue-length based policy, respectively.}\label{greedy_Q_upload_fig}
\end{figure}
Second, we consider a file upload scenario where each user transmitting a file with finite size to
the base station in a rateless manner. Let $\mathcal{T}_i$ be the $i$-th user's completion time of
the file upload session for a file of size $f_i$.  Define the average upload rate for the $i$-th
user as $\frac{f_i}{\mathcal{T}_i}$. We can measure the performance of each policy for this
scenario by evaluating the utility function at the average upload rate. Figure
\ref{greedy_Q_upload_fig} demonstrates the utility difference of the greedy and the queue-based
policy for different file sizes. We can observe that for small file sizes the greedy policy
outperforms the queue-based policy significantly, and this difference decreases by increasing the
file size. We can interpret this behavior as follows. The files are first buffered into the queues
based on the queue lengths and the weighted $\alpha$-fair utility, while the queues are emptied by
a max-weight scheduler. Once the files are all buffered in the queues, the queues are empties with
the same rate which is not fair because it does not give any priority to the users based on their
utility. For larger file size, the duration for which the entire file is emptied into the queue is
negligible compared to the total transmission time, and the average upload rate converges to a
near-optimal rate.

\section{Conclusion}
We addressed the problem of optimal resource allocation in a fading multiple access channel from an
information theoretic point of view. We formulated the problem as a utility maximization problem
for a more general class of utility functions.

We considered several different scenarios. First, we considered the problem of optimal rate
allocation in a non-fading channel. We presented the notion of approximate projection for the
gradient projection method to solve the rate allocation problem in polynomial time in the number of
users.

Second, we studied rate and power allocation in a fading channel with known channel statistics. In
this case, the optimal rate and power allocation policies are obtained by greedily maximizing a
properly defined linear utility function. If for the fading channel power control and channel
statistics are not available, the greedy policy is not optimal for nonlinear utility functions.
However, we showed that its performance in terms of the utility is not arbitrarily worse compared
to the optimal policy, by bounding their performance difference. The provided bound tends to zero
as the channel variations become small or the utility function behaves more linearly.

The greedy policy may itself be computationally expensive. A computationally efficient algorithm
can be employed to allocate rates close to the ones allocated by the greedy policy. Two different
rate allocation policies are presented which only take one iteration of the gradient projection
method with approximate projection at each time slot. It is shown that these policies track the
greedy policy within a neighborhood which is characterized by average speed of fading as well as
fading speed in the worst case.

\appendices
\section{Algorithm for finding a violated constraint}\label{appendix_Rate_splitting}

In this section, we present an alternative algorithm based on rate-splitting idea to identify a
violated constraint for an infeasible point. For a feasible point, the algorithm provides
information for decoding by successive cancellation. We first introduce some definitions.

\begin{definition}\label{config}
   The quadruple $(M, \bs P, \bs R, N_0)$ is called a \textit{configuration} for an $M$-user
    multiple-access channel, where $\bs R = (R_1, \ldots, R_M)$ is the rate tuple, $\bs P = (P_1, \ldots, P_M)$ represents the
    received power and $N_0$ is the noise variance. For any given configuration, the \emph{elevation}, $\bs \delta \in \mathbb R^M$, is
    defined as the unique vector satisfying
    \begin{equation}\label{elevation}
                R_i = C(P_i, N_0+ \delta_i), \quad i =1,\ldots, M.
    \end{equation}
\end{definition}

    Intuitively, we can think of message $i$ as rectangles of height $P_i$, raised above the noise
    level by $\delta_i$. In fact, $\delta_i$ is the amount of additional Gaussian interference that
    message $i$ can tolerate.
    Note that if the rate vector corresponding to a configuration is feasible its elevation vector is non-negative. However, that is
    not sufficient for feasibility check.

\begin{figure}
  \centering
  \includegraphics[width=.6\textwidth]{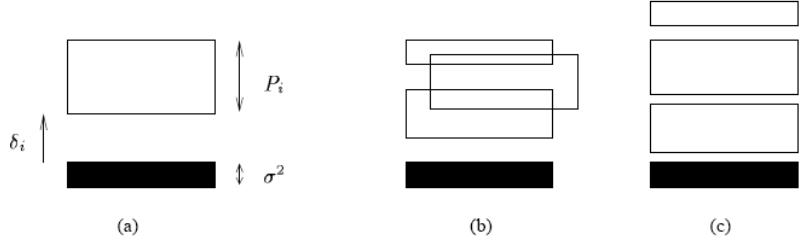}\\
  \caption{Graphical representation of messages over multi-access channel \cite{Urbanke}.}\label{ratesplit_fig}
\end{figure}

\begin{definition}\label{codable}
    The configuration $(M, \bs P, \bs R, N_0)$ is \emph{single-user codable}, if after possible re-indexing,
    \begin{equation}\label{su_codable}
        \delta_{i+1} \geq \delta_i + P_i,   i = 0, 1 , \dots, M-1,
    \end{equation}
    where we have defined $\delta_0 = P_0 = 0$ for convention.
\end{definition}
    By the graphical representation described earlier, a configuration is single-user codable if
    the none of the messages are overlapping. Figure \ref{ratesplit_fig}(a) gives an example of graphical representing
    for a message with power $P_i$ and elevation $\delta_i$. Figures \ref{ratesplit_fig}(b) and
    \ref{ratesplit_fig}(c) illustrate overlapping and non-overlapping configurations, respectively.

\begin{definition}
    The quadruple $(m, \bs p, \bs r, N_0)$ is a \textit{spin-off} of $(M, \bs P, \bs R, N_0)$ if there exists a
    surjective mapping $\phi : \{1,\ldots, m\} \rightarrow \{1,\ldots, M\}$ such that for all $i \in \{1,\ldots,
    M\}$ we have

\begin{eqnarray}
   P_i &\geq& \sum_{j \in \phi^{-1}(i)} p_j, \nonumber \\
   R_i &\leq& \sum_{j \in \phi^{-1}(i)} r_j. \nonumber
 \end{eqnarray}
    where $\phi^{-1}(i)$ is the set of all $j \in \{1,\ldots,m\}$ that map into $i$ by means of $\phi$.
\end{definition}

\begin{definition}\label{hyper_user}
       A \textit{hyper-user} with power $\bar{P}$, rate $\bar{R}$, is obtained by merging $d$ actual users
        with powers $(P_{i_1}, \ldots, P_{i_d})$ and rates $(R_{i_1}, \ldots, R_{i_d})$, i.e,
\begin{equation}\label{hyper}
   \bar{P} = \sum_{k=1}^d P_{i_k}, \quad
          \bar{R} =\sum_{k=1}^d R_{i_k}.
\end{equation}

\end{definition}

\begin{theorem}\label{rate_splitting_thm}
For any $M$-user achievable configuration $(M, \bs P, \bs R, N_0)$, there exists a spin-off $(m,
\bs p , \bs r, N_0)$ which is single user codable.
\end{theorem}
\begin{proof}
See Theorem 1 of \cite{Urbanke}.
\end{proof}
Here, we give a brief sketch of the proof to give intuition about the algorithm. The proof is by
induction on $M$. For a given configuration, if none of the messages are overlapping then the
spin-off is trivially equal to the configuration. Otherwise, merge the two overlapping users into a
\emph{hyper-user} of rate and power equal the sum rate and sum power of the overlapping users,
respectively. Now the problem is reduced to rate splitting for $(M-1)$ users. This proof suggests a
recursive algorithm for rate-splitting that gives the actual spin-off for a given configuration.

It follows directly from the proof of Proposition \ref{rate_splitting_thm} that this recursive
algorithm gives a single-user codable spin-off for an achievable configuration. If the
configuration is not achievable, then the algorithm encounters a hyper-user with negative
elevation. At this point the algorithm terminates. Suppose that hyper-user has rate $\bar R$ and
power $\bar P$. Negative elevation is equivalent to the following
$$\bar R > C(\bar P, N_0).$$
Hence, by Definition \ref{hyper_user} we have,
$$\sum_{i \in S} R_i > C(\sum_{i \in S} P_i, N_0).$$
where $S = \{i_1,\ldots,i_d\} \subseteq \mathcal M$. Therefore, a hyper-user with negative
elevation leads us to a violated constraint in the initial configuration.

\begin{theorem}\label{ratesplit_complexity}
The presented algorithm runs in $O(M^2\log M)$ time, where $M$ is the number of users.
\end{theorem}
\begin{proof}
The computational complexity of the algorithm can be computed as follows. The algorithm terminates
after at most $M$ recursions. At each recursion, all the elevations corresponding to a
configuration with at most $M$ hyper-users are computed in $O(M)$ time. It takes $O(M\log M)$ time
to sort the elevation in an increasing order. Once the users are sorted by their elevation, a
hyper-user with negative elevation could be found in $O(1)$ time, or two if such a hyper-user does
not exists it takes $O(M)$ time to find two overlapping hyper-users. In the case that there are no
overlapping users and all the elevations are non-negative the input configuration is achievable,
and the algorithm terminates with no violated constraint. Hence, computational complexity of each
recursion is $O(M)+O(M\log M)+O(M) = O(M\log M)$. Therefore, the algorithm runs in $O(M^2\log M)$
time.
\end{proof}

\section{Proof of Lemma \ref{region_chebyshev}}\label{chebyshev_pf}
First, consider the following lemmas. Lemma \ref{jensen_diff1} bounds Jensen's difference of a
random variable for a concave function. The upper bound is characterized in terms of the variance
of the random variable.

        \begin{lemma} \label{jensen_diff1}
            Let $f: \mathbb{R} \rightarrow \mathbb{R}_+$ be concave and twice differentiable.
            Let $X$ be a random variable with variance $\sigma_X^2$. Then,
            \begin{equation}\label{jensen_diff}
                f(\mathbb{E}[X]) - \mathbb{E}[f(X)] \leq \sqrt{2M\sigma_X^2 f(\mathbb{E}[X])} - \frac{\sigma_X^2
                M}{2},
            \end{equation}
            where $M$ be an upper-bound on $|f''(x)|$.
        \end{lemma}

        \begin{proof}
        Pick any $0 < \epsilon \leq 1$. By Chebyshev's inequality we have
        \begin{equation}\label{chebyshev}
           \textbf{Pr}\left(|X-\mathbb{E}(X)| > c\right) \leq \epsilon,
        \end{equation}
        where $c = \frac{\sigma_X}{\sqrt{\epsilon}}$. Therefore, we have
        \begin{eqnarray}
          \mathbb{E}[f(X)] &=& \mathbb{E}\left[f(X) \Big| |X-\mathbb{E}(X)| \leq c\right] \textbf{Pr}\big(|X-\mathbb{E}(X)| \leq c\big) \nonumber \\
          &+& \mathbb{E}\left[f(X) \Big| |X-\mathbb{E}(X)| > c\right] \textbf{Pr}\big(|X-\mathbb{E}(X)| > c\big) \nonumber \\
          &\geq& (1-\epsilon)  \mathbb{E}\left[f(X) \Big| |X-\mathbb{E}(X)| \leq c\right] \nonumber \\
          &\geq& \frac{1-\epsilon}{2}\bigg(f\big(\mathbb{E}[X]+ c\big) + f\big(\mathbb{E}[X]- c\big) \bigg)\nonumber \\
          &=& (1-\epsilon) f(\mathbb{E}[X]) + \frac{1-\epsilon}{4} c^2 (f''(\xi_1)+f''(\xi_2)),
          \end{eqnarray}
            where the first inequality follows from non-negativity of $f$, and the second and the
            second inequality follows from concavity of $f$. The scalars $\xi_1 \in
            \big[\mathbb{E}[X], \mathbb{E}[X]+c\big]$ and $\xi_2 \in
            \big[\mathbb{E}[X]-c,\mathbb{E}[X]\big]$ are given by Taylor's theorem.

            Given the above relation, for any $\epsilon > 0$ we have
            \begin{equation}\label{jen2}
                f(\mathbb{E}[X]) - \mathbb{E}[f(X)] \leq \frac{1-\epsilon}{2\epsilon} \sigma_X^2
                M + \epsilon f(\mathbb{E}[X]).
            \end{equation}

            The right-hand side is minimized for
            \begin{equation}\label{epsilon_min}
                \epsilon^* = \min\left\{ \left(\frac{\sigma_X^2 M}{2
                f(\mathbb{E}[X])}\right)^{\frac{1}{2}},1\right\}.
            \end{equation}

            By substituting $\epsilon^*$ in (\ref{jen2}), the desired result follows immediately.

        \end{proof}

        We next provide an upper bound on variance of $Y= \log(1+X)$ proportional to the variance
        of $X$.

        \begin{lemma}\label{var_Y}
            Let $X > 0$ be a random variable with mean $\bar{X}$ and variance $\sigma_X^2$, and $Y
            = \log(1+X)$ then variance of $Y$ is upper-bounded as
            \begin{equation}\label{Y_var}
                \sigma_Y^2 \leq \sigma_X^2\left(1+ \left[(1+\bar{X})(\sqrt{2 \log(1+\bar{X}}) -
               \frac{\sigma_X}{2})\right]^2\right).
            \end{equation}
        \end{lemma}
        \begin{proof}
            Let $\mathbb E(Y) = \log(1+\hat{X})$ for some $\hat{X} < \bar{X}$. By invoking the mean value theorem, we have
            \begin{eqnarray}\label{varY_eqn1}
              \sigma_Y^2 &=& \mathbb{E}\bigg[\Big(\log(1+X)-\log(1+\hat{X})\Big)^2\bigg] \nonumber \\
               &=& \mathbb{E}\bigg[\Big(\frac{1}{1+\tilde{X}}(X-\hat{X})\Big)^2\bigg] \nonumber \\
               &\leq& \mathbb{E}\Big[\big(X-\hat{X}\big)^2\Big],
            \end{eqnarray}
            where $\tilde{X}$ is a non-negative random variable.

            On the other hand, by employing lemma \ref{jensen_diff1} with $f(x) = \log(1+x)$, we can write
            \begin{equation}\label{expect_bnd}
                \mathbb{E}\big[\log(1+X)\big] \geq \log(1+\bar{X}) - \sqrt{2\sigma_X^2 \log(1+\bar{X})} + \frac{\sigma_X^2
                }{2}.
            \end{equation}

            Hence,
            \begin{eqnarray}
              \bar{X} \geq \hat{X} &=& \exp\left\{  \mathbb{E}[\log(1+X)] \right\} - 1 \nonumber \\
               &\geq& \exp\left\{  \log(1+\bar{X})  - \sqrt{2\sigma_X^2 \log(1+\bar{X})} + \frac{\sigma_X^2 }{2} \right\} - 1 \nonumber \\
               &\geq&  \bar{X} - \sigma_X(1+\bar{X})(\sqrt{2 \log(1+\bar{X}}) - \frac{\sigma_X
               }{2}),
            \end{eqnarray}
            where the first inequality is by (\ref{expect_bnd}), and the second relation can be
            verified after some straightforward manipulation. By combining (\ref{varY_eqn1}) and
            (\ref{expect_bnd}) the variance of $Y$ can be bounded as follows
            \begin{eqnarray}
              \sigma_Y^2 &\leq& \mathbb{E}[(X-\hat{X})^2] \nonumber \\
               &\leq& \mathbb{E}\left[\left(X-\bar{X} + \sigma_X(1+\bar{X})(\sqrt{2 \log(1+\bar{X}}) - \frac{\sigma_X
               }{2})\right)^2\right] \nonumber \\
               &=& \sigma_X^2\left(1+ \left[(1+\bar{X})(\sqrt{2 \log(1+\bar{X}}) -
               \frac{\sigma_X}{2})\right]^2\right).
            \end{eqnarray}
        \end{proof}

Now we provide the proof for Lemma \ref{region_chebyshev}. Define random variable $Y_S$ as the
following:

\begin{equation}\label{Y_S}
 Y_S = \frac{1}{2} \log(1+\sum_{i \in S}\frac{H_i P_i}{N_0}), \quad \textrm{for all}\ S \subseteq \mathcal M = \{1, \ldots, M\}.
\end{equation}

The facet defining constraints of $C_g(\bs P, \bs H)$ and $C_a(\bs P)$ are of the form of $\sum_{i
\in S}R_i \leq Y_S$ and $\sum_{i \in S}R_i \leq \mathbb{E}[Y_S]$, respectively. Therefore, by
Definition \ref{Hausdorff_def}, we have $d_H \left(C_g(\bs{P},\bs{H}), C_a(\bs{P}) \right) \leq
\delta$ if and only if $|Y_S - \mathbb{E}[Y_S]| \leq \delta$, for all $S \subseteq \mathcal M$.
Thus, we can write

\begin{eqnarray}\label{chebyshev_pf_eqn}
   \textbf{\textrm{Pr}} \Big\{ d_H \left(C_g(\bs{P},\bs{H}), C_a(\bs{P}) \right) > \delta \Big\} &=& \textbf{Pr}\Big\{ \max_{S} \big|Y_S - \mathbb{E}[Y_S]\big| > \delta \Big\}
   \nonumber \\
   &\leq & \sum_{S \subseteq \mathcal M} \textbf{Pr}\Big\{ \big|Y_S - \mathbb{E}[Y_S]\big| > \delta \Big\}
   \nonumber \\
   &\leq& \frac{1}{\delta^2}\sum_{S \subseteq \mathcal M}\sigma^2_{Y_S}.
\end{eqnarray}
where the first inequality is obtained by union bound, and the second relation is by applying
Chebyshev's inequality. On the other hand, $\sigma^2_{Y_S}$ can be bounded from above by employing
Lemma \ref{var_Y}, i.e.,
\begin{equation}\label{sigma_Y}
    \sigma^2_{Y_S} \leq \frac{\sigma_{Z_S}^2}{4}\left(1+ \left[(1+\bar{Z}_S)(\sqrt{2 \log(1+\bar{Z}_S)} -
    \frac{\sigma_{Z_S}}{2})\right]^2\right),
\end{equation}
where
$$ \bar{Z}_S = \mathbb{E}\Big[ \sum_{i \in S}\frac{H_i P_i}{N_0}\Big] = \sum_{i \in S} \Gamma_i \bar{H_i} = \bs \Gamma'_S \bs{\bar{H}}, $$
$$ \sigma_{Z_S}^2 = \textrm{var}\Big( \sum_{i \in S}\frac{H_i P_i}{N_0}\Big) = \sum_{(i,j) \in S^2} \Gamma_i \Gamma_j \textrm{cov}(H_i, Hj) =\bs \Gamma_S' K \bs \Gamma_S. $$

The desired result is concluded by substituting $\bar{Z_S}$ and $\sigma_{Z_S}^{2}$ in
(\ref{sigma_Y}) and combing the result with (\ref{chebyshev_pf_eqn}).    $\qquad \blacksquare$

\section{Proof of Lemma \ref{opt_dist}}\label{opt_dist_pf}

Let us first state and prove a useful lemma which asserts that Euclidean expansion of a capacity
region by $\delta$ contains its expansion by relaxing its constraints by $\delta$.

\begin{lemma}\label{expansion_incl}
Let $C_1$ be a capacity region with \emph{polymatroid} structure, i.e.,
            \begin{equation}\label{polymatroid}
                C_1 = \bigg\{ \bs R \in \mathbb{R}^M_+: \sum_{i \in S} R_i \leq f(S),\
                \textrm{for all}\ S \subseteq \mathcal M \bigg\},
            \end{equation}
where $f(S)$ is a nondecreasing submodular function. Also, let $C_2$ be an \emph{expansion} of
$C_1$ by $\delta$ as defined in Definition \ref{expansion_def}. Then, for all $\bs R \in C_2$,
there exists some $\bs R' \in C_1$ such that $\|\bs R - \bs R'\| \leq \delta$.
\end{lemma}
\begin{proof}
By Definition \ref{submodular_def}, it is straightforward to show that $C_2$ is also a polymatroid,
i.e.,
            \begin{equation}\label{polymatroid2}
                C_2 = \bigg\{ \bs R \in \mathbb{R}^{M}_+: \sum_{i \in S} R_i \leq g(S) = f(S) + \delta,\
                \textrm{for all}\ S \subseteq {\mathcal M}  \bigg\},
            \end{equation}
where $g(S)$ is a submodular function. By convexity of $C_2$, we just need to prove the claim for
the vertices of $C_2$. Let $\bs R \in \R^{M}$ be a vertex of $C_2$. The polymatroid structure of
$C_2$ implies that $\bs R$ is generated by an ordered subset of ${\mathcal M}$ (see Theorem 2.1 of
\cite{polymatroid_vertex}). Hence, there is some $k \in \mathcal{M}$ such that $ R_k = f(\{k\}) +
\delta$. Consider the following construction for $\bs R'$:
                 \begin{equation}\label{R'_R}
                   R'_i = \left\{ \begin{array}{ll}
                   R_i - \delta, & \textrm{$i = k$}\\
                   R_i, & \textrm{otherwise.}
                   \end{array} \right.
                 \end{equation}

By construction, $\bs R'$ is  in a $\delta$-neighborhood of $\bs R$. So we just need to show that
$R'$ is feasible in $C_1$.  First, let us consider the sets $S$ that contain $k$. We have
                \begin{equation}\label{k_in_S}
                    \sum_{i \in S} R'_i = \sum_{i \in S} R_i - \delta \leq f(S).
                \end{equation}
                Second, consider the case that $k \notin S$.
                \begin{eqnarray}
                  \sum_{i \in S} R'_i &=& \sum_{i \in S \cup \{k\}} R'_i - R_k + \delta \nonumber \\
                   &\leq& f(S \cup \{k\})  + \delta - R_k \nonumber \\
                   &\leq& f(S) + f(\{k\}) + \delta - R_k \nonumber \\
                   &=& f(S), \nonumber
                \end{eqnarray}
                where the first inequality comes from (\ref{k_in_S}), and the second inequality is
                true by submodularity of the function $f(\cdot)$. This completes the proof.
\end{proof}

\begin{proof}[of Lemma \ref{opt_dist}]
                Without loss of generality assume that $u(\bs R_2^*) \geq u(\bs R_1^*)$. By Lemma
                \ref{expansion_incl}, there exists some $\bs R \in C_a(\bs P)$ such that $\|\bs
                R_2^* - \bs R\| \leq \delta$. Moreover, we can always choose $\bs R$ to be on the
                boundary so that $\|\bs R\| \geq D_\delta$, where $D_\delta$ is defined in
                (\ref{diameter}). Therefore, by Assumption \ref{assumption_u2}(a) and the fact that
                $u(\bs R_2^*) \geq u(\bs R_1^*) \geq u(\bs R)$, we have
                \begin{equation}\label{opt_dist1}
                    u(\bs R_2^*) - u(\bs R) = |u(\bs R_2^*) - u(\bs R)| \leq B\|\bs R_2^* - \bs R\| \leq B
                    \delta.
                \end{equation}

                Now suppose that $\|\bs R_1^* - \bs R\| > (\frac{B}{A} \delta)^{\frac{1}{2}}$. By Assumption \ref{assumption_u2}(b) we
                can write
                \begin{equation}\label{opt_dist2}
                    u(\bs R_1^*) - u(\bs R) = |u(\bs R_1^*) - u(\bs R)| \geq A\|\bs R_1^* - \bs R\|^2 >  B \delta.
                \end{equation}
                By subtracting (\ref{opt_dist1}) from (\ref{opt_dist2}) we obtain $u(\bs R_2^*) <
                u(\bs R_1^*)$ which is a contradiction. Therefore, $\|\bs R_1^* - \bs R\| \leq
                (\frac{B}{A} \delta)^{\frac{1}{2}}$, and the desired result follows immediately by invoking the
                triangle inequality.
\end{proof}

\bibliographystyle{unsrt}
\bibliography{MAC}

\end{document}